\documentclass[3p,sort&compress]{elsarticle}

\usepackage{amssymb}    % Mathematical symbols
\usepackage{amsmath}  % More options for mathematics
\usepackage{epstopdf}   % Convert eps to pdf
\usepackage[separate-uncertainty=true]{siunitx}   % Proper formatting of units in math mode
\usepackage{color}      % Supports text color if needed
\usepackage{soul}       % https://ctan.org/pkg/soul
\usepackage{lmodern}    % Loading fonts
\usepackage{hyperref}   % To insert clickable references/urls
\usepackage{listings}   % To input code in the text
\usepackage{amsmath}
\usepackage{amsmath}
\usepackage{amssymb}
\usepackage{graphicx}
\usepackage[lofdepth,lotdepth]{subfig}
\usepackage{epstopdf}
\usepackage{booktabs}
\usepackage{braket}
\usepackage[utf8]{inputenc}
\usepackage[T1]{fontenc}
\usepackage{lipsum}
\journal{supervisor}
\begin{document}

\begin{frontmatter}
    \title{Quantum relaxation in a system of harmonic oscillators with time-dependent coupling}
\author[add1]{F. B. Lustosa}
\ead{chicolustosa@cbpf.br}
\author[add2, add3]{S. Colin}
\ead{samuel.colin@cyu.fr}
\author[add4]{S. E. Perez Bergliaffa}
\ead{sepbergliaffa@gmail.com}

\address[add1]{Centro Brasileiro de Pesquisas Físicas, Rua Dr. Xavier Sigaud 150,
Urca - CEP: 22290-180 - Rio de Janeiro-RJ,  Brazil}
\address[add2]{Theiss Research, 7411 Eads Avenue, La Jolla, CA 92037, United States}
\address[add3]{Laboratoire de Physique Th\'eorique et Mod\'elisation, CNRS Unit\'e 8089,  
CY Cergy Paris Universit\'e, 95302 Cergy-Pontoise cedex, France}
\address[add4]{Departamento de Física Teórica, Instituto de Física, Universidade do Estado do Rio de Janeiro, R. São
Francisco Xavier, 524, Maracanã - CEP: 20559-900 - Rio de Janeiro-RJ, Brazil.}

    \begin{abstract}
In the context of the de Broglie-Bohm pilot wave theory, numerical simulations for simple systems have shown that %quantum 
states that are initially 
out of quantum equilibrium - thus 
violating the Born rule -
%- nonequilibrium - 
usually relax 
over time
to the expected $|\psi|^2$ distribution  on a coarse-grained level. We analyze the relaxation of nonequilibrium initial distributions for a system of coupled one-dimensional harmonic oscillators in which the coupling depends explicitly on time through numerical simulations, 
focusing in the influence of
%taking into account 
different parameters such as the number of modes, the coarse-graining length and the coupling constant. We show that in general the system studied here tends to equilibrium, but the 
relaxation 
can be retarded depending on the values of the parameters, particularly to the one related to the strength of 
%giving special focus to 
the interaction. Possible implications on the detection of relic nonequilibrium systems are discussed. 
    \end{abstract}
    
\end{frontmatter}

%% How to make a heading and divide the documents into different sections
%
%\twocolumn
%
\section{Introduction}

The de Broglie-Bohm pilot-wave theory \cite{bib:Bohm1951a, bib:Bohm1951b, bib:Bohm1993, bib:Holland1995} allows the existence of states that violate the Born rule and provides a mechanism through which quantum probabilities can emerge \cite{bib:Valentini1990, bib:Valentini1991, bib:Valentini1992, bib:Valentini2001, bib:Valentini2005, bib:Towler2012, bib:Valentini2019}. In this context, the known results of quantum theory reflect the current equilibrium state of most physical systems, but systems out of the so-called quantum equilibrium may also exist \cite{bib:Valentini1990, bib:Valentini1991, bib:Valentini1992, bib:Valentini1996, bib:Valentini2001, bib:Valentini2005, bib:Valentini2001uq, bib:Valentini2004ep, bib:Valentini2004xv, bib:Valentini2005st, bib:Valentini2007, bib:Valentini2008, bib:Valentini2010,bib:Colin2013, bib:Underwood2015, bib:Underwood2017, bib:Kandhadai2016}. Simulations 
of the evolution 
of two-dimensional systems have shown that relaxation to equilibrium occurs - on a coarse-grained level - approximately exponentially for most states formed by a  sufficient number of
superposed modes  \cite{bib:Valentini2005, bib:Towler2012, bib:Colin2011, bib:Colin2013, bib:Abraham2014}. Some systems may have been out of quantum equilibrium in the very early universe \cite{bib:Valentini1996,bib:Valentini2001, bib:Valentini2005st, bib:Valentini2007, bib:Valentini2008, bib:Valentini2010, bib:Valentini2015}. If so, an observable effect on the cosmic microwave background (CMB) may be present \cite{bib:Valentini2010, bib:Valentini2015, bib:Colin2013,bib:Colin2015, bib:Colin2016, bib:Vitenti2019}. Out-of-equilibrium relic systems that decoupled at a very early time have a different spectrum from that predicted by the Copenhagen interpretation of quantum mechanics (which will be referred as orthodox quantum theory from now on) that could be detected today \cite{bib:Valentini2008,  bib:Underwood2015, bib:Underwood2017}. 

The current state of equilibrium of the systems analyzed in laboratories is often %related 
linked
to their ``violent astrophysical history'' \cite{bib:Valentini2005, bib:Valentini2010} and, on the discussion about non-equilibrium relic systems, interactions between particles are said to be ``likely to cause rapid relaxation'' \cite{bib:Underwood2015}. However, only one-particle systems or unentangled modes have been considered in simulations (with the exception of \cite{bib:Underwood2015}, where a time independent coupled scalar system was studied). Although
%the link between relaxation and interactions seems natural
it is natural to expect a correlation between relaxation and interactions, there is no evidence yet 
%there is no \textit{a priori} motive to expect that 
supporting the hypothesis that 
stronger interactions would induce faster relaxation.
%without a proper numerical analysis. 
We begin to address the problem of interacting systems out of quantum equilibrium in this paper. 

In previous works the relaxation to quantum equilibrium was analysed for two main types of systems: the two dimensional potential well with one particle \cite{bib:Valentini2005, bib:Towler2012} and the two dimensional harmonic oscillators \cite{bib:Abraham2014, bib:Underwood2018}\footnote{A model of with a scalar field was studied in \cite{bib:Colin2015} but it was ultimately shown to be analyzable in terms of harmonic oscillators. The model of coupled scalar fields studied in \cite{bib:Underwood2015} was limited to a two level system and focused on the transference of nonequilibrium from one field to the other.}. In those simulations the motivation can be viewed as two-folded: i) it is important to show that relaxation does take place efficiently to justify the fact that all systems observed so far seem to be in equilibrium; ii) but it is also important to provide a way through which nonequilibrium can be observed and so there is interest in finding systems that \textit{have not} relaxed completely yet or even at all. It is still an open question if \textit{in general} most systems do relax efficiently, but available numerical evidence suggests that this is the case. Numerical results also suggest that there is a number of parameters that can delay or even prevent relaxation. Only further study of different models and a wider 
%variation 
examination of the
parameters space can provide a more complete picture of the relaxation process. However, since the evidence up to now indicates simultaneously rapid relaxation for most systems and the %possibility 
possible existence
of residues of nonequilibrium that last a long time \cite{bib:Abraham2014, bib:Kandhadai2016, bib:Underwood2018}, it is interesting to explore realistic scenarios where incomplete relaxation could leave an observable imprint,  cosmology being a natural place to look for it \cite{bib:Valentini2008, bib:Valentini2010, bib:Colin2013,  bib:Valentini2014, bib:Valentini2015, bib:Colin2015, bib:Underwood2015, bib:Colin2016, bib:Underwood2017}.

This paper analyzes a system with a time dependent coupling, specifically a system composed of two one dimensional harmonic oscillators with a product type coupling. The main 
%objective 
goal
of our analysis is to test the effect of the coupling on the relaxation timescale. We will also test the effect of 
%varying 
the variation of other relevant parameters, whose influence was %that were also present
tested 
in previous simulations. 

This work is organized as follows. Before going into the details of our model, the development and the current state of the quantum relaxation picture 
are reviewed in Section \ref{qr},
where a detailed summary and discussion of the numerical results are presented. Section \ref{exactsol} is devoted to the derivation of the exact wave function of the system of coupled oscillators. We study in Section 
\ref{numres}
the behaviour of the evolution of our model using numerical techniques. We first define our general setup, showing in subsection 4.1 that different sets of initial phases or combination of quantum states give significantly different relaxation timescales. In subsections 4.2 and 4.3 we study the averaged effect of the variation of parameters averaged over a set of randomly chosen initial states. In Section \ref{conc} we conclude with a discussion of our results and the possible significance for the relaxation of systems in cosmological scenarios. 

\section{Quantum relaxation and numerical simulations}
\label{qr}

The de Broglie-Bohm theory describes the deterministic motion of individual 
quantum systems, as opposed to the ensemble description furnished by Copenhagen interpretation. Given an initial configuration $q(0)$ and the wave function $\psi(q,t)$ it is possible to determine the exact trajectory of the system by solving de Broglie's guidance equation
\begin{equation} \label{eq:1}
    \dot{q}(t) = \frac{j(q,t)}{|\psi(q,t)|^{2}}
\end{equation}
where $\psi(q,t)$ obeys a Schrödinger equation and $j(q,t)$ is the current associated with the continuity equation:
\begin{equation} \label{eq:2}
    \frac{\partial |\psi(q,t)|^{2}}{\partial t} + \partial_q \cdot j(q,t) = 0.
\end{equation}
The solution to the Schrödinger equation determines the wave function at all times and the corresponding guidance equation determines the evolution of $q(t)$. In any experiment the initial position cannot be known with  exact precision, so it is useful to consider a distribution $\rho(q,0)$ of initial positions 
%with the same wave function 
that also obeys a continuity equation given by
\begin{equation} \label{eq:3}
    \frac{\partial\rho(q,t)}{\partial t} + \partial_q \cdot (\rho(q,t)\dot{q}(t)) = 0.
\end{equation}
Since equations (\ref{eq:2}) and (\ref{eq:3}) are identical, if %initially 
$\rho(q,0) = |\psi(q,0)|^{2}$, then the Born rule is valid for all values of $t$. However, 
%the de Broglie-Bohm theory does not impose the initial conditions on physical systems and 
there is no reason in principle for the distribution
of initial positions to be equal to $|\psi(q,0)|^{2}$. In fact, the evolution %consists solely 
is governed by 
the Schrödinger and guidance equations, being the initial positions and distributions contingencies of each set up to  which the theory is to be applied. This naturally implies the question: if the de Broglie-Bohm theory is valid why do we always observe systems that comply to 
the
the Born rule? 

There are currently two main approaches to this issue: one based on a `Bohmian' view of the de Broglie-Bohm theory that explains the Born rule through a `typicality' argument for the initial configuration of the whole universe\footnote{For a detailed review and critique of this approach we refer to \cite{bib:Valentini2019} and references therein.} \cite{bib:Durr1992, bib:Durr2009, bib:Goldstein2017, bib:Tumulka2018} and the `quantum relaxation' picture \cite{bib:Valentini1990, bib:Valentini1991}, where nonequilibrium distributions evolve to the expected $|\psi|^2$ value 
%on a coarse-grained level. 
in a manner analogous to thermal relaxation. This later approach was developed in terms of a coarse-grained $H$-theorem in configuration space analogous to Gibb's classical coarse-grained $H$-theorem in phase space. Assuming that the initial distribution has no microstructure, it was shown in \cite{bib:Valentini1990, bib:Valentini1991} that a coarse-grained $H$-function defined by %
\begin{equation} \label{eq:4}
\bar{H}(t) = \int dq \bar{\rho}(q,t) \ln \left(\frac{\bar{\rho}(q,t)}{\overline{|\psi(q,t)|^2}} \right),
\end{equation}
always decreases in time. Furthermore, numerical simulations
were carried out for initially out-of-equilibrium two-dimensional
systems with different numbers of energy modes superposed, and most of them show an approximately exponential decay to equilibrium of the form $\bar{H}(t) \approx \bar{H}(0)\exp(-t/\tau)$  \cite{bib:Valentini2005, bib:Towler2012, bib:Colin2013, bib:Colin2011, bib:Abraham2014}, where $\tau$ is the relaxation timescale.

One of the goals of these simulations was to find a general correlation between the timescale $\tau$ for relaxation and the different numerical parameters of the systems, specifically, the number of superposed modes $M$ and the coarse-graining length $\epsilon$. A first attempt in this direction was made in \cite{bib:Valentini2005}, where 
%the system of 
a particle in a two-dimensional potential well was first studied. In that work, a relation of the type $\tau \propto 1/\epsilon$ was found to compare favourably with the numerical simulations. 
%A test of 
The dependence
%relation 
of $\tau$ with the number of modes was not %obtained 
studied due to computational limitations. Using the same theoretical model of the particle in the box but with a more complete and slightly different numerical analysis, in \cite{bib:Towler2012} different number of modes where considered and the results pointed to a relation of the type $\tau \propto M^{-1}$. The same form of exponential decay was confirmed but 
%there was 
no definite dependence on the coarse-graining length was found. 

Another system that has been extensively studied through numerical simulations is the two-dimensional harmonic oscillator \cite{bib:Contopoulos2012, bib:Colin2013, bib:Abraham2014, bib:Kandhadai2016} which is of special interest since it can be shown to be mathematically equivalent to the uncoupled mode of a real scalar field \cite{bib:Valentini2007, bib:Valentini2008, bib:Colin2013, bib:Colin2015, bib:Underwood2015, bib:Colin2016, bib:Kandhadai2016}. The approximate exponential decay was still observed for most $\bar{H}(t)$ calculated \cite{bib:Colin2013, bib:Abraham2014} but the possibility of saturation of relaxation for small $M$ (with some particular choices of initial phases) was also observed \cite{bib:Contopoulos2012, bib:Abraham2014, bib:Kandhadai2016}. 

The number of superposed energy states is equal to the number of nodes in the wave function and those, in turn, have been related to the chaotic nature of the trajectories in the de Broglie-Bohm theory \cite{bib:Frisk1997, bib:Wu1999, bib:Wisniacki2005, bib:Efthymiopoulos2006, bib:Efthymiopoulos2007, bib:Efthymiopoulos2017}. This chaos has been generally considered as fundamental to the relaxation process \cite{bib:Valentini2005, bib:Towler2012, bib:Efthymiopoulos2017}, implying that 
%. That would mean that
more complex and chaotic systems (with larger $M$) should relax faster then simpler systems, which % that 
could have slower relaxation or even no relaxation at all \cite{bib:Abraham2014, bib:Kandhadai2016, bib:Underwood2018}. The test of this hypothesis through direct calculation of the $\bar{H}(t)$ 
%function for ever growing number of parameters  (superposed modes, number of test particles, evolution time and different initial states) 
is computationally cumbersome 
due to the large number of parameters involved (superposed modes, number of test particles, evolution time and different initial states),
so another approach, first introduced in \cite{bib:Contopoulos2012} and based on the character of the trajectories, has been used for the study of long-time relaxation \cite{bib:Abraham2014, bib:Kandhadai2016}. It led 
%and there was 
to evidence that even for some initial states with $M = 25$ complete quantum equilibrium could not be reached. More recently, using yet another numerical approach, Underwood \cite{bib:Underwood2018} has shown that it is possible that some states of extreme nonequilibrium can completely avoid relaxation. These results indicate that the presence of nodes can induce faster acceleration for most states but it is still an open question (that has begun to be answered in \cite{bib:Underwood2018}) if some other factors can cause systems to stay out of quantum equilibrium even for large numbers of $M$.

Aside from the number of modes, it is important to point out some other %characteristics 
features
of the initial quantum states studied so far in the simulations. In general, the states considered were constructed as evenly weighted superposed energy modes with randomly assigned phases $\theta_{i,j}$. As was pointed out in \cite{bib:Towler2012} different initial phases correspond to different initial positions of the nodes of the wave function and they can result in different relaxation timescales as was already observed for the case $M=4$ in that work. However, \cite{bib:Towler2012} claims that this would only considerably affect the process of decay to equilibrium for a low number of nodes and consider the mean value of $\tau$ obtained with 6 different sets of phases for the larger values of $M$ because the average distribution of nodes will be similar to one another. On the other hand, one could imagine that some sets of phases cause some confined distribution of nodes and hence delay the relaxation process even for systems with many modes superposed. A more detailed examination of this feature was made on Section 5 of \cite{bib:Abraham2014}, where it was shown that some choices of phases cause complete relaxation even for $M = 4$, while others could generate a fixed residue in the $\bar{H}(t)$ function for $M=25$. The conclusion of that work is that relaxation is more `likely' for larger values $M$, but a precise measure of this `likelihood' was not obtained. In \cite{bib:Kandhadai2016} some choices of phases also caused significantly different physical behaviours for the trajectories. These results indicate that when considering the process of quantum relaxation it is important to consider a wide set of initial phases and that it is still not clear how the positions of nodes affect the timescale of the decay. It is possible that further investigation in the role of vorticity \cite{bib:Efthymiopoulos2017, bib:Underwood2018} on the confinement of trajectories can shed some light into that question.

Another fundamental characteristic of the initial states considered in \cite{bib:Valentini2005, bib:Towler2012, bib:Colin2013, bib:Abraham2014, bib:Kandhadai2016} is the choice of basis for the energy states. In those works the Hamiltonian was written in a Cartesian 2D basis that provides a simple set of uncoupled trajectory equations for the 2 coordinates. That also leads to what can be called a fine-tunning \cite{bib:Underwood2018} when choosing a symmetric combination of quantum numbers with a definite total energy, because different sets of states could have been chosen with the same energy (i.e. a set $[\psi_{00}, \psi_{01}, \psi_{10}, \psi_{11}]$ has the same total energy as $[\psi_{00}, \psi_{01}, \psi_{10}, \psi_{02}]$). In our work we try to avoid such fine-tunning by randomly creating a set of initial states that are not necessarily symmetric (in the sense that we can have a set that contains $\psi_{02}$ but not $\psi_{20}$). Since the exact initial quantum state of a system cannot be known directly, we take a statistical approach and try to learn general information assuming only that the total energy of the state is known, allowing to study systems with any number of superposed modes instead of the symmetric $M=4, 9, 16, 25$ etc. studied so far.

\section{Exact solution for the time-dependent coupled Schr\"{o}dinger equation}
\label{exactsol}

We will consider a system of two interacting harmonic oscillators with the same masses and fundamental frequencies, and a product interaction with a time dependent coupling of the form $k(t) = \beta t$. The coupled Hamiltonian of the whole system is given by
\begin{equation}
H(x_{a}, x_{b}, t) = \frac{p_{a}^{2}}{2m} + \frac{m\omega^{2}x_{a}^{2}}{2} +  \frac{p_{b}^{2}}{2m} +  \frac{m\omega^{2}x_{b}^{2}}{2} + m k(t) x_{a}x_{b}.
\end{equation}
With the coordinate transformations
\begin{align}
x_{a} = \frac{1}{\sqrt{2}}\left(x_{1} + x_{2}\right), \label{tc1}\\
x_{b} = \frac{1}{\sqrt{2}}\left(x_{1} - x_{2}\right),
\label{tc2}
\end{align}
the 
Hamiltonian is expressed as the addition of two non-interacting oscillators with time-dependent frequency:
\begin{equation}
H(x_{1}, x_{2}, t) = \sum_{r=1}^2H_{r} = \sum_{r=1}^2\left(\frac{p_{r}^{2}}{2m} +  m\Omega_{r}^2x_{r}^{2}\right), 
\end{equation}
where $\Omega_{r}^{2} = (\omega^{2} \pm \beta t)$, the plus (minus) sign applies when $r=1(r=2)$. 
%Upon quantization,  
The corresponding Schr\"{o}dinger equation can be written as
\begin{equation}
    \imath\frac{\partial \Psi}{\partial t} = \sum_{r=1}^2\left( -\frac{1}{2m}\frac{\partial^2}{\partial x_r^2} + \frac{m}{2}\Omega_r^2x_r^2\right)\Psi.
\end{equation} 
This equation coincides with the one  describing an uncoupled field mode in an expanding background
\cite{bib:Colin2013}, but with different time dependence in the mass and frequency.
Hence, we shall follow the derivation of the exact wave function of the system presented there. 
The first step is the expansion of the  wave function at $t = 0$ in terms of the initial eigenstates $\Phi_{n_r}(x_r)$ of the  Hamiltonian 
and write it as
\begin{equation}
\Psi(x_{1}, x_{2}, 0) = \sum_{n_{1}, n_{2}} c_{n_{1}, n_{2}}(0) \Phi_{n_1}(x_{1}) \Phi_{n_2}(x_{2}).
\end{equation} 
The $\Phi_{n_r}(x_r)$ states will then evolve according to one-dimensional Schrödinger equations and the solution at time $t$ can then be written in the original coordinates as the product
\begin{equation}
\Psi(x_{a}, x_{b}, t) = \sum_{n_{1}, n_{2}} c_{n_{1}, n_{2}}(0) \psi_{n_1}(x_{1}(x_{a},x_{b}),t) \psi_{n_2}(x_{2}(x_{a},x_{b}),t)
\label{totalpsi}
\end{equation} 
of the independent solutions $\psi_{n_r}(x_{r},t)$ with initial conditions $\psi_{n_r}(x_{r},0) = \Phi_{n_r}(x_r)$. These, in turn, can be obtained with the methods presented  in \cite{bib:Ji1995} and have the form
\begin{equation}
\psi_{n_{r}} (x_{r}, t) = \frac{1}{\sqrt{2^{n_r}n_r !}}F_r(t)G_r(x_r,t), 
\end{equation}
where
\begin{equation}
      F_r(t) = \frac{\omega_{r,I}^{\frac{1}{4}}\exp\left(-\imath(n_r+\frac{1}{2})\int_0^t {dt}'\frac{\omega_{r,I}}{(m g_{r,-}(t))}\right)}{(\pi g_{r,-}(t))^{\frac{1}{4}}},
\end{equation}
\begin{equation}
   \begin{gathered}
     G_r(x_r,t) = \exp\left(-\frac{x_r^2}{2g_{r,-}(t)}(\omega_{r,I}+\imath g_{r,0}(t))\right) \\
        \times \mathcal{H}_{n_r}\left(\sqrt{\frac{\omega_{r,I}}{g_{r,-}(t)}}\right),
\end{gathered} 
\end{equation}where $\omega_{r,I} = \sqrt{g_{r,+}(t)g_{r,-}(t)-g_{r,0}^2(t)}$ is a constant of motion, and the $\mathcal{H}_{n_r}$ are  the usual Hermite polynomials. The functions $g_{r,\pm}(t)$ and $g_{r,0}(t)$ are solutions of the
following system of differential equations
\begin{align}
\dot{g}_{j,-} = -2\frac{g_{j,0}}{m}, \label{deq1}\\
\dot{g}_{j,0} = m\Omega_{j}^{2}g_{j,-} - \frac{g_{j,+}}{m}, \label{deq2}\\
\dot{g}_{j,+} = 2m\Omega_{j}^{2}g_{j,0},
\end{align}
with initial conditions given by $g_{j,-}(0) = 1/m$, $g_{j,0}(0) = 0$ e $g_{j,+}(0) = m\Omega_{j}(0)^{2}$. The most general solution for $g_{r,-}(t)$ is given by \cite{bib:Ji1995}
\begin{equation}
g_{r,-} = c_{r,1}f_{r,1}^{2} + c_{r,2}f_{r,1}f_{r,2} + c_{r,3}f_{r,2}^{2},
\end{equation}
where $f_{r,1}$ and $f_{r,2}$  are two independent solutions of the equation
\begin{equation}
\ddot{f} + \Omega_{r}^{2}f = 0.
\end{equation}
With the solutions of this equation for $r=1,2$, $g_{r,-}(t)$ can be written
in terms of the constants $c_{r,j}$ ($j=1,2,3$), and substituting the resultant expression in the differential equations \eqref{deq1} and \eqref{deq2}
we find the other functions $g_{r,0}$ and $g_{r,+}$. Finally, with the initial conditions we fix the constants and construct the full wave function.

For the case considered here ($\Omega_r=\sqrt{\omega^2 \pm \beta t}$), the four solutions are given by
\begin{align}
f_{r, 1} = \frac{\Omega_{r}}{\beta^{1/3}}\mathit{J}_{1/3}\left(\frac{2}{3}\frac{\Omega_{r}^{3}}{\beta}\right), \\
f_{r, 2} = \frac{\Omega_{r}}{\beta^{1/3}}\mathit{Y}_{1/3}\left(\frac{2}{3}\frac{\Omega_{r}^{3}}{\beta}\right),
\end{align}
and the functions $\mathit{J}_{1/3}$ and $\mathit{Y}_{1/3}$ are the Bessel functions of the first and second kinds, respectively. This solutions are valid only in the interval  $t \in [0,\omega^2/\beta]$ because for $r=2$ the Bessel functions are ill-defined for $t > \omega^2/\beta$. 
%This is actually the interval of physical interest since classical trajectories seem to diverge as $t \rightarrow \omega^2/\beta$, 
In fact, we have checked that the classical trajectories oscillate up to 
$t\approx \omega^2/\beta$ and then 
start to diverge, thus signalling that the interaction adopted is no longer consistent with 
treating the whole system as coupled oscillators  \footnote{Actually, a one-dimensional problem similar to the one studied here 
%the case with $\Omega_1$
was treated using standard quantum mechanics in \cite{bib:Ji1995, bib:Agarwal1991} and at a time equivalent to $t = \omega^2/\beta$ a transition in the interaction, defined by $\beta t \rightarrow \beta$, was introduced.}.

We are now in a position to construct trajectory equations. Before doing so it is important to notice that we are actually interested in the trajectories $x_a(t)$ and $x_b(t)$, however, transformations (15) and (16) are linear as are the guidance equations. We will then use the strategy of applying the transformations of coordinates for the initial conditions $(x_a(0), x_b(0)) \rightarrow (x_1(0), x_2(0))$, then solve numerically the guidance equations for the transformed coordinates and transform the results back to the original system for each step at which we wish to calculate the $H$-function. That being said, the guidance equations for the transformed coordinates $(x_1, x_2)$ can be written as
\begin{align}
    \dot{x}_1 = \operatorname{Im}\left(\frac{1}{\Psi}\frac{\partial \Psi}{\partial x_1}\right), \\
    \dot{x}_2 = \operatorname{Im}\left(\frac{1}{\Psi}\frac{\partial \Psi}{\partial x_2}\right).
\end{align}
Upon replacement of the solution of the Schrödinger equation, the above equations are a set of coupled first order differential equations that can be solved numerically, given any initial positions $(x_a(0), x_b(0))$. Assuming a distribution of initial positions that is different from $|\Psi(x_a, x_b, 0)|^2$ and calculating the trajectories for each initial condition we can simulate the evolution of any distribution $P(x_a, x_b, t)$ that will allow us to calculate the $\bar{H}(t)$ function given in Eq.\eqref{eq:4}.

\section{Numerical results}
\label{numres}

In the last section an exact wave function for the Hamiltonian of the system of coupled one-dimensional harmonic oscillators was presented. Here we give the details of the  numerical simulations
that 
were implemented to evaluate the evolution of $\bar{H}$ for different values of the parameters. We shall see that the variation of such values has different impacts on the timescales for relaxation to quantum equilibrium.

\subsection{Numerical setup}

In order to calculate $\bar{H}(t)$ we simulate the evolution of a distribution of trajectories $\rho(x_a, x_b, t)$ by numerically solving a number 2N of guidance equations with different initial conditions. We distribute the initial positions using a random Gaussian generator for each calculation of the $H$-function. The initial positions are distributed in a box in configuration space in the area of the support of the wave function, which consists of the intervals $[x_a = -5...5, x_b= -5..,5]$. 

After generating the initial distribution, we use the transformation of coordinates given in Eqs.\eqref{tc1} and \eqref{tc2}
to obtain the transformed initial positions. Using a Runge-Kutta method of order 8 \cite{bib:Hairer1993} we then proceed to solve the guidance equations given in Eqs. (31) and (32), storing the solutions of each equation after transforming it back to the original coordinate system:
\begin{align}
    (x_a(0), x_b(0)) \rightarrow (x_1(0), x_2(0)), \\
(x_1(0), x_2(0)) \rightarrow (x_1(t), x_2(t)), \\
(x_1(t), x_2(t))  \rightarrow (x_a(t), x_b(t)).
\end{align}
The Runge-Kutta method is applied for each trajectory initially with an absolute error tolerance of $10^{-5}$ and then with one ten times smaller. If the difference between each of the results for $x_1$ and $x_2$ is larger than the cutoff value given by $\delta = 0.0025$, the calculation is performed again with a smaller error tolerance until the minimum value of $10^{-15}$ is attained.

\clearpage
\begin{figure}
  \begin{minipage}{\linewidth}
  %\hfill
  \includegraphics[width=.3\linewidth]{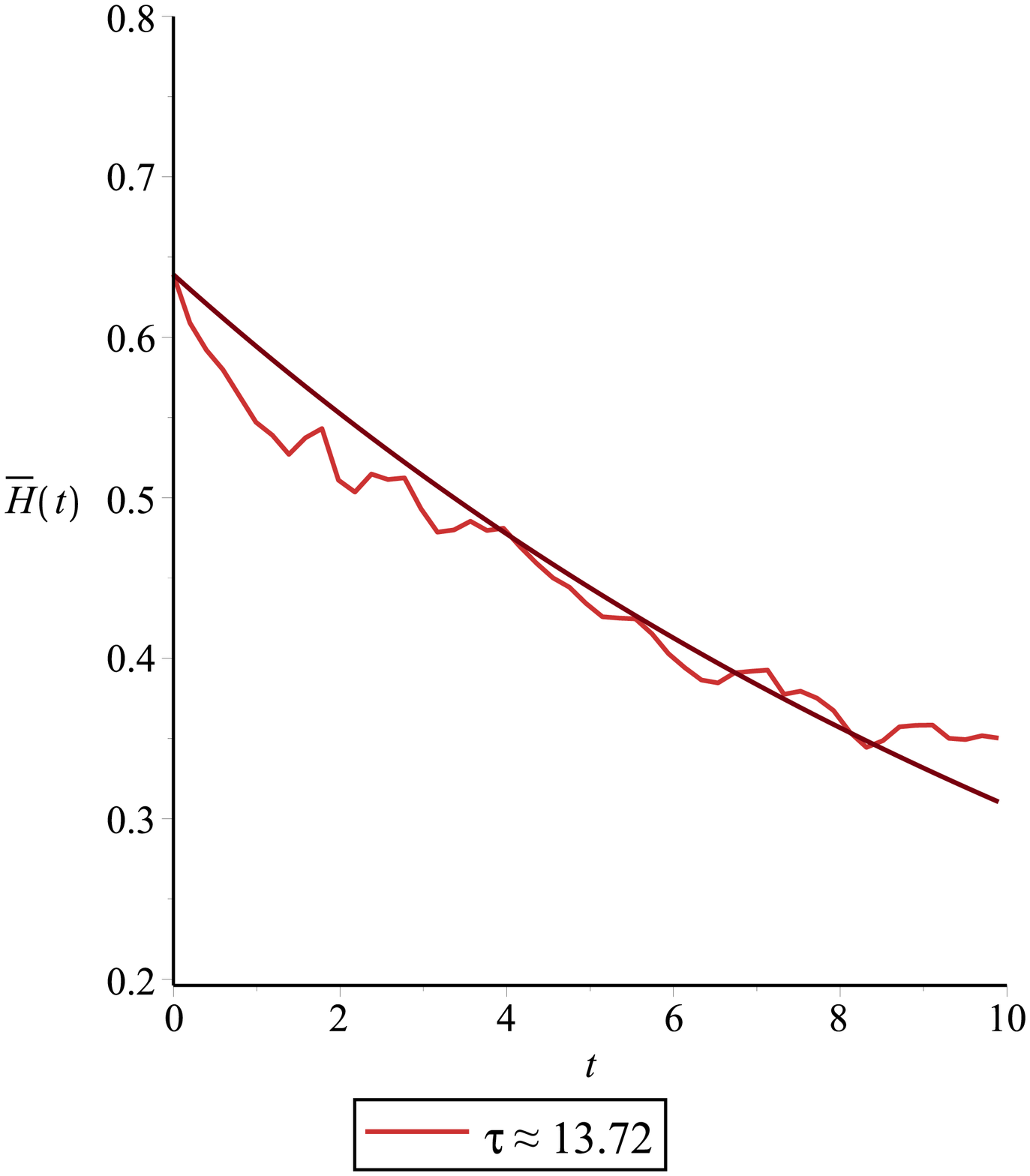}
  %\hfill
  \includegraphics[width=.3\linewidth]{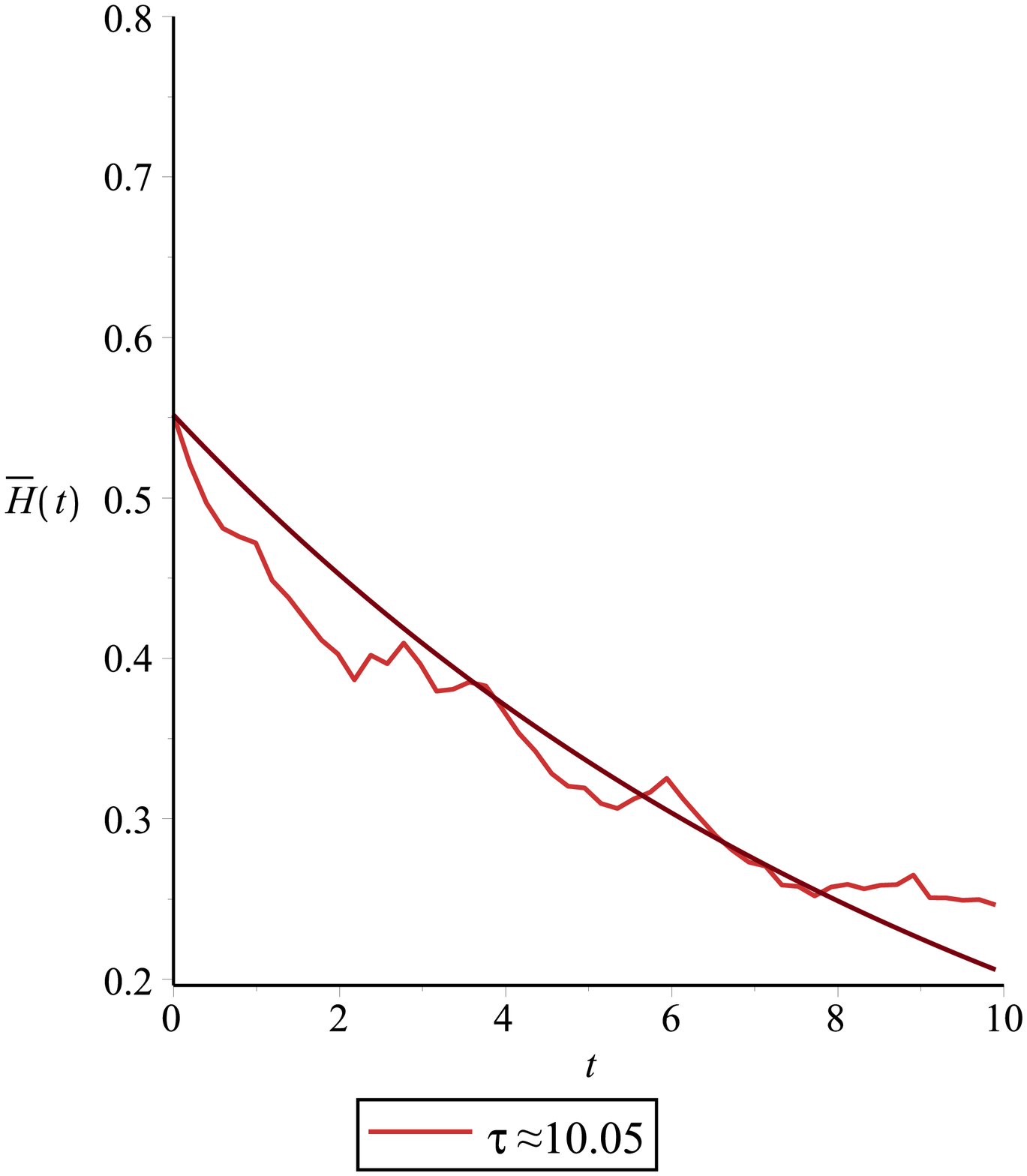}
  %\hfill
  \includegraphics[width=.3\linewidth]{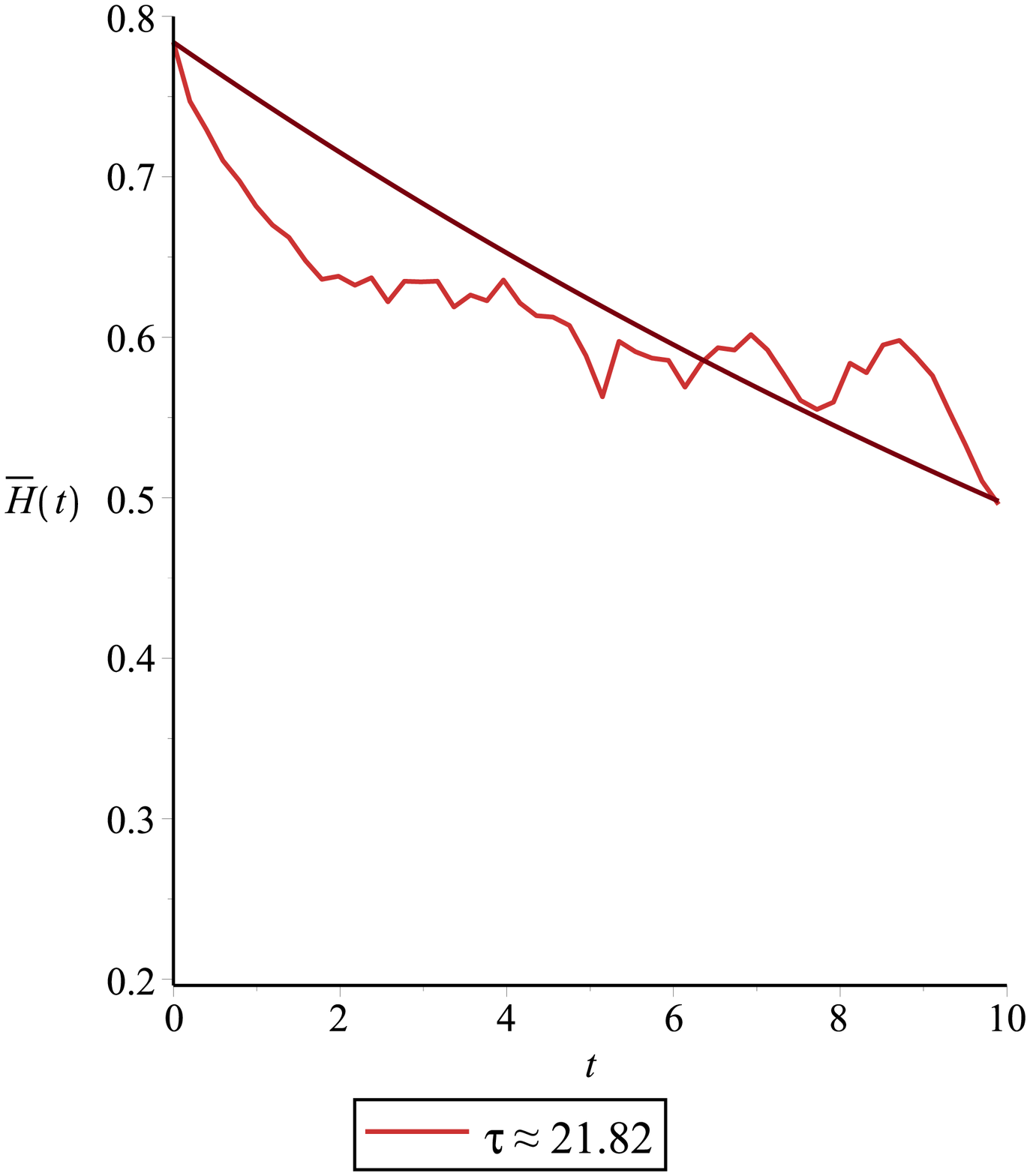}
  %\hfill
%  \includegraphics[width=.3\linewidth]{example-image-a}%
  \end{minipage}%
   \caption{$\bar{H}(t)$ functions for $M=15$ and different sets of initial phases.}
   \label{m15}
\end{figure}

Using the wave function described in the previous section and having a distribution of positions for a number of timesteps we calculate the coarse-grained values of both functions $|\psi|^{2}$ and $\rho$. The coarse-graining is 
performed
by dividing the two-dimensional box in configuration space in cells of side $\epsilon$ and then numerically calculating the integral of the trajectories that are inside each cell for the distribution. The coarse-grained values of both the wave function and the distribution are then substituted in the equation
\begin{equation}
    \bar{H}(t) = \int \int \bar{\rho}(x_a, x_b, t)\ln{\frac{\bar{\rho}(x_a, x_b, t)}{\overline{|\Psi(x_a,x_b,t)|^2}}}dx_a dx_b,
\end{equation}
which is also numerically integrated to give values of $\bar{H}(t)$ at each of the timesteps. 

In each of the steps outlined above a number of parameters are involved. For simplicity, we choose $m = \omega = 1$. The initial wavefunctions are defined up to a phase factor that is given by 
\begin{equation}
c_{n_1 , n_2}(0)=\frac{\exp{2\pi\imath\theta_{n_1,n_2}}}{\sqrt{M}},
\end{equation}
where $M$ is the number of modes superposed. Besides choosing the phase and the number of modes there is the choice of the quantum states that are superposed, defined by the quantum numbers $n_1$ and $n_2$
(see Eq.\eqref{totalpsi}). For each value of $M$,
$(n_1,n_2) $ and
the phases $\theta_{n_1,n_2}$ were randomly generated.

With the initial wave function defined, we have to set the number $N$ of pairs 
of trajectories that  will be calculated, and the number of coarse-graining cells that define the
value of $\epsilon$. Finally, the coupling constant $\beta$ has to be chosen to determine the 
%scale 
strength
of the interaction.
A range of choices of $\beta$ will be scanned to evaluate the effect of the interaction in the evolution of $\bar{H}(t)$. All of these parameters can potentially affect the time scale of relaxation towards quantum equilibrium so they will be examined separately.

\subsection{Choice of initial quantum states}

Quantum trajectories tend to be highly chaotic in configuration space and they are naturally sensitive to the initial conditions. As already shown in previous works \cite{bib:Abraham2014, bib:Kandhadai2016}, different choices of initial phases can lead to different relaxation timescales and might even result in residual nonequilibrium after a long time. 
The plots of Figure \ref{m15} show that this is also the case in the system under scrutiny. The  $\bar{H}(t)$ functions in such plots were calculated for a wavefunction with $M=15$, and the plots also show the best fit of the form $H_{0}\exp{(-t/\tau)}$, from which the corresponding values for the relaxation timescales were obtained  \footnote{For the plots in this subsections, we chose $\epsilon = 0.2$, $N= 2 \cdot 10^5$ and $\beta = 0.1$.}. For the case of 15 modes superposed only one combination of quantum numbers is possible, so the only difference between each calculation is the choice of phases $\theta_{n_1,n_2}$. The  values of $\tau$ are of the same order of magnitude but clearly depend of the choice of phases. We have confirmed this dependence with further simulations using different sets of values of  $\theta_{n_1,n_2}$. Such a dependence 
might be connected to the position of the nodes that correspond to each energy mode, but since they are also one of the motives for the chaotic nature, their precise  effect on the relaxation is probably random. Hence, we will employ in the next subsections ten sets of randomly generated phase values for each calculation of the $\bar{H}(t)$ function.

The relaxation process 
can also be influenced by the quantum numbers. In most of the previous works a conservative choice was made: only what we will refer to as symmetric superpositions of states with values of $M$ that have an integer square root were considered. For example, in those works a wave function with $M=4$ was composed by the quantum states $[\psi_{0,0},\psi_{1,0},\psi_{0,1},\psi_{1,1}]$, thus not including the states $\psi_{2,0}$ and $\psi_{0,2}$. In our simulations, besides randomly generating the phases for each state we also assign random quantum numbers to each choice of $M$.
\begin{figure}[!ht]

  \begin{minipage}{\linewidth}
  %\hfill
  \includegraphics[width=.3\linewidth]{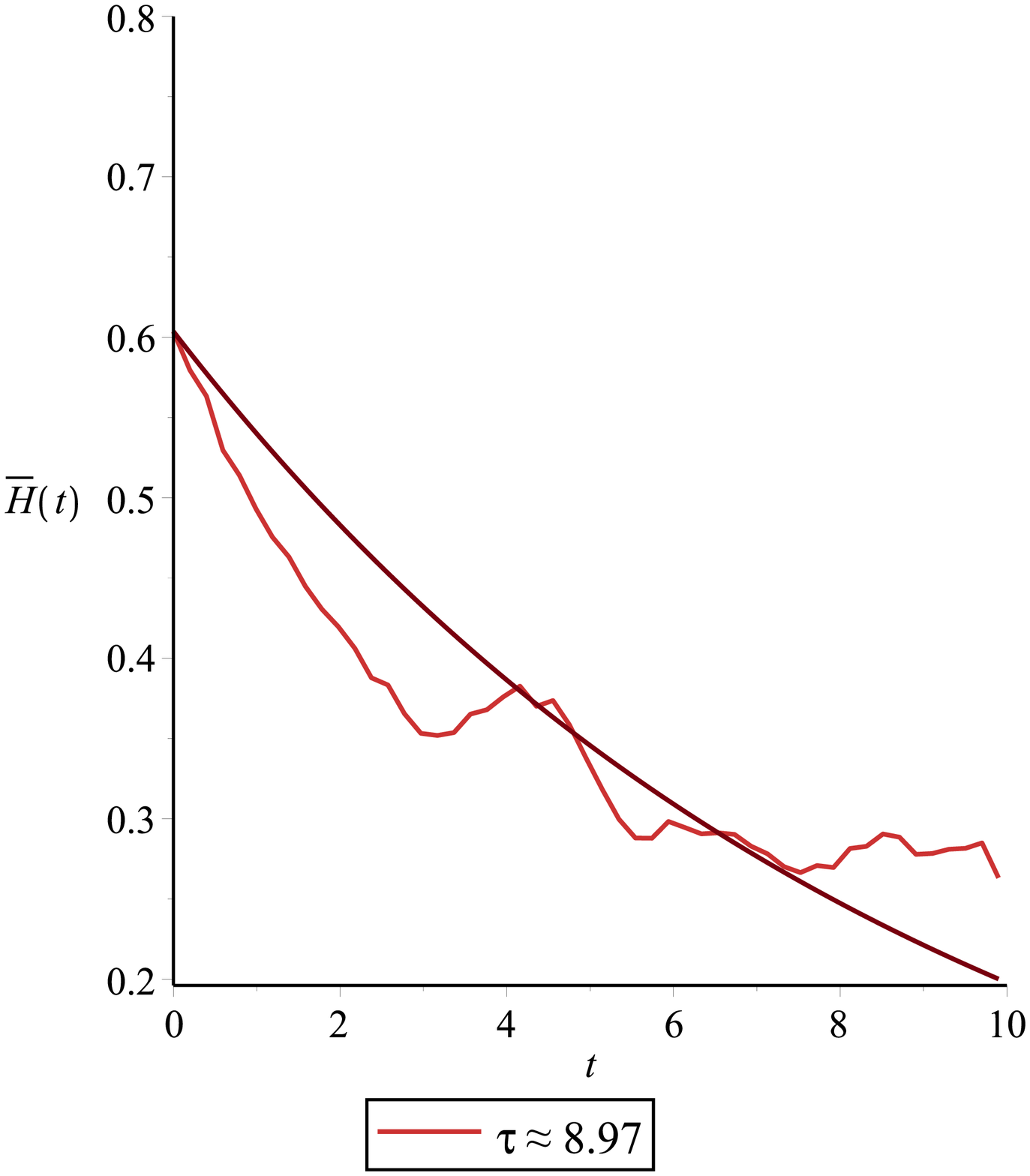}
  %\hfill
  \includegraphics[width=.3\linewidth]{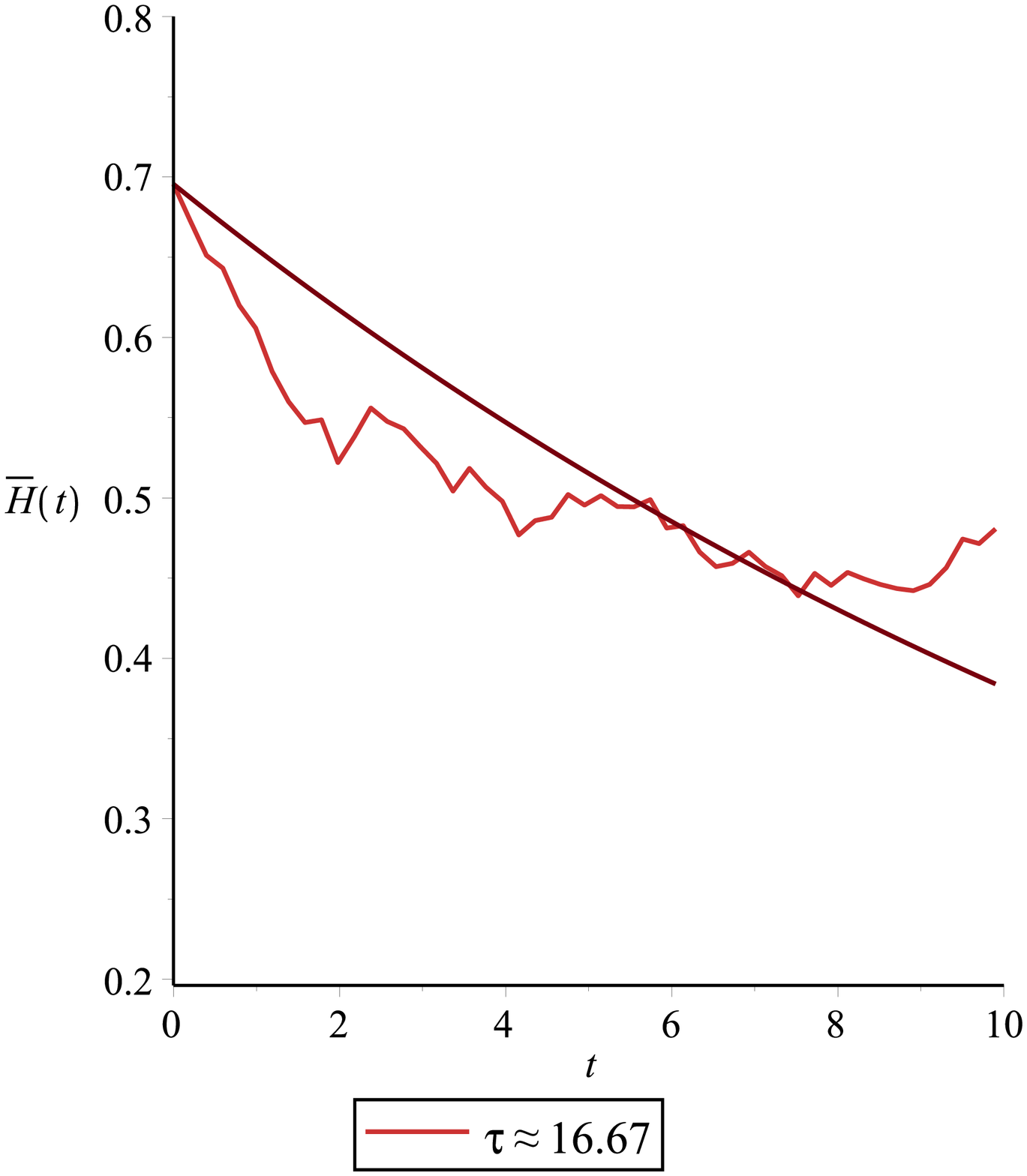}
  %\hfill
  \includegraphics[width=.3\linewidth]{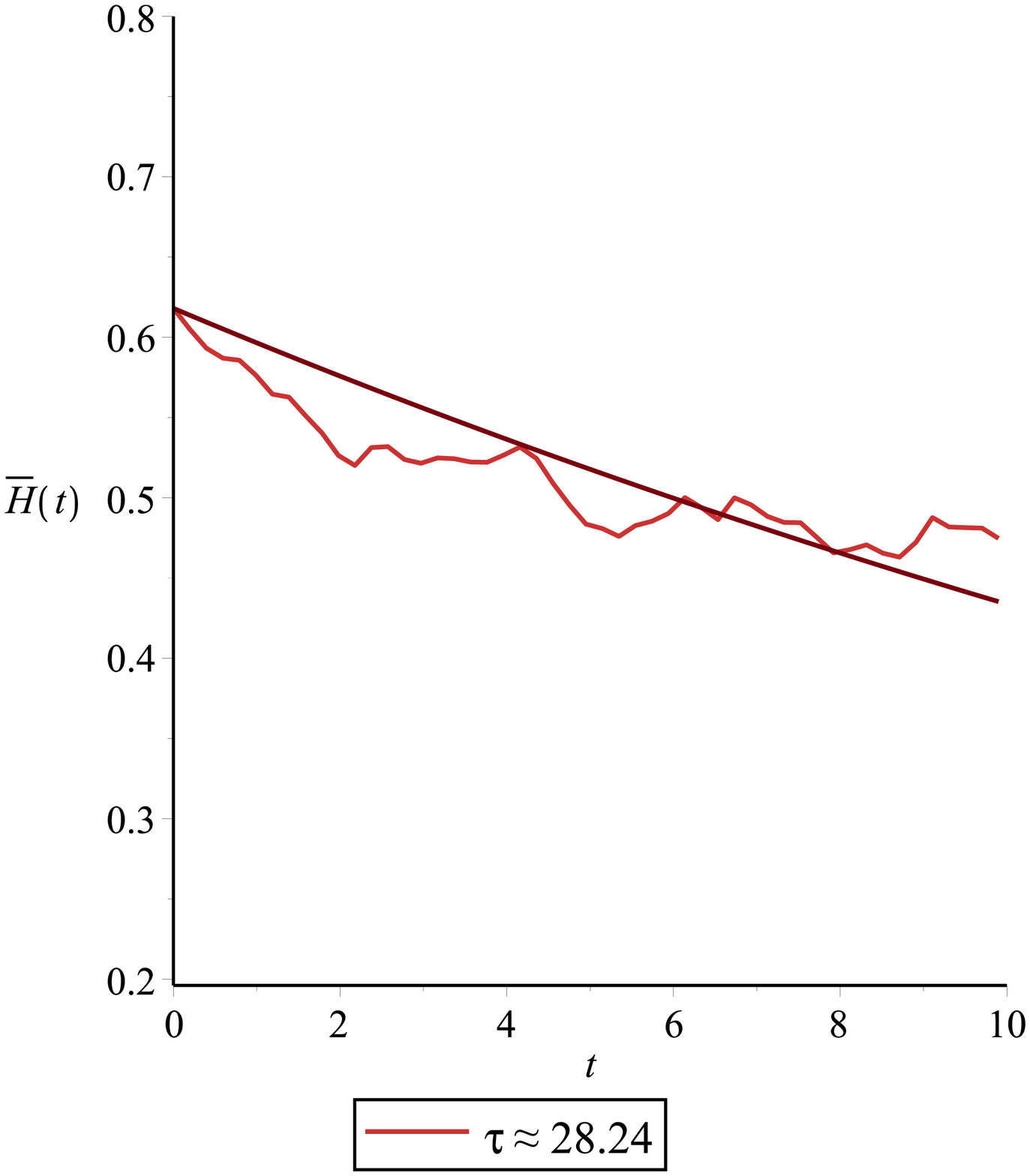}
  %\hfill
%  \includegraphics[width=.3\linewidth]{example-image-a}%
  \end{minipage}%
   \caption{$\bar{H}(t)$ functions for $M=9$, different sets of initial phases and different combinations of quantum numbers.}
\label{m9}
\end{figure}
In Figure \ref{m9} we show three different plots for $\bar{H}(t)$ and their corresponding best fits with the timescales $\tau$ associated to 
three different presets of phases and quantum numbers for the case $M=9$. The states considered have, respectively, the following quantum numbers:
\begin{align}
\left[\psi_{0,0},\psi_{0,1},\psi_{1,0},\psi_{0,2},\psi_{1,1},\psi_{2,0},\psi_{2,1},\psi_{0,3},\psi_{1,2}\right], \\
 \left[\psi_{0,0},\psi_{0,1},\psi_{1,0},\psi_{0,2},\psi_{1,1},\psi_{2,0},\psi_{1,2},\psi_{3,0},\psi_{0,3}\right], \\
\left[\psi_{0,0},\psi_{0,1},\psi_{1,0},\psi_{0,2},\psi_{1,1},\psi_{2,0},\psi_{2,1},\psi_{1,2},\psi_{0,3}\right].
\end{align}
The timescales again have the same order of magnitude but they  differ significantly for each combination.
Although it is not possible through our analysis to separate 
the effects of different phases and quantum numbers, the numerical evidence obtained so far, both  in our work and in the aforementioned previous ones, suggests that such choices have an apparently 
random effect on the relaxation timescales. 
Hence, in the following ten presets of phases and quantum numbers 
will be considered 
for each simulation
in the analysis of the evolution of the $\bar{H}(t)$ function. The values for the $H$-functions will then be averaged over this ten choices and those averages will then be used to calculate the relaxation timescale through the best fit function $H_{0}\exp{(-t/\tau)}$.

\subsection{Coarse-graining length and the number of modes}

The coarse graining length $\epsilon$ defines the area of the cells over which the densities and wave functions are calculated. Previously the simulations were done using a \textit{bactracking} algorithm to ensure that a fixed number of trajectories were on each cell. In our simulations we go forward in time and as we vary the value of $\epsilon$ we also change the value of calculated trajectories to avoid having cells with a small number of particles. The backtracking mechanism comes with the problem that at each desired time step the whole trajectory needs to be calculated, while in our case we just need to store the positions of the particles at each of the 50 times at which we calculate the $H$-function.

\begin{figure}
\centering
\subfloat{
  \includegraphics[width=0.8\linewidth, height=0.2\textheight]{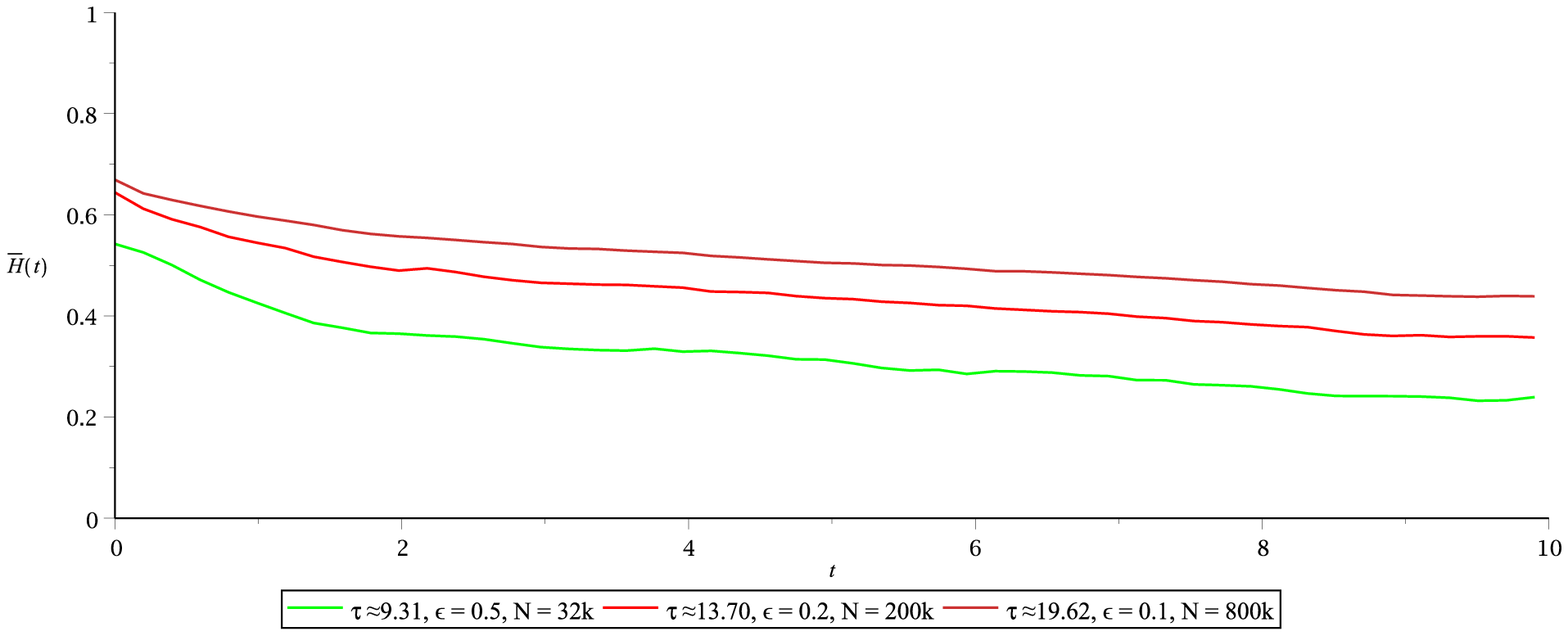}}  
\par
\subfloat{
  \includegraphics[width=0.8\linewidth, height=0.2\textheight]{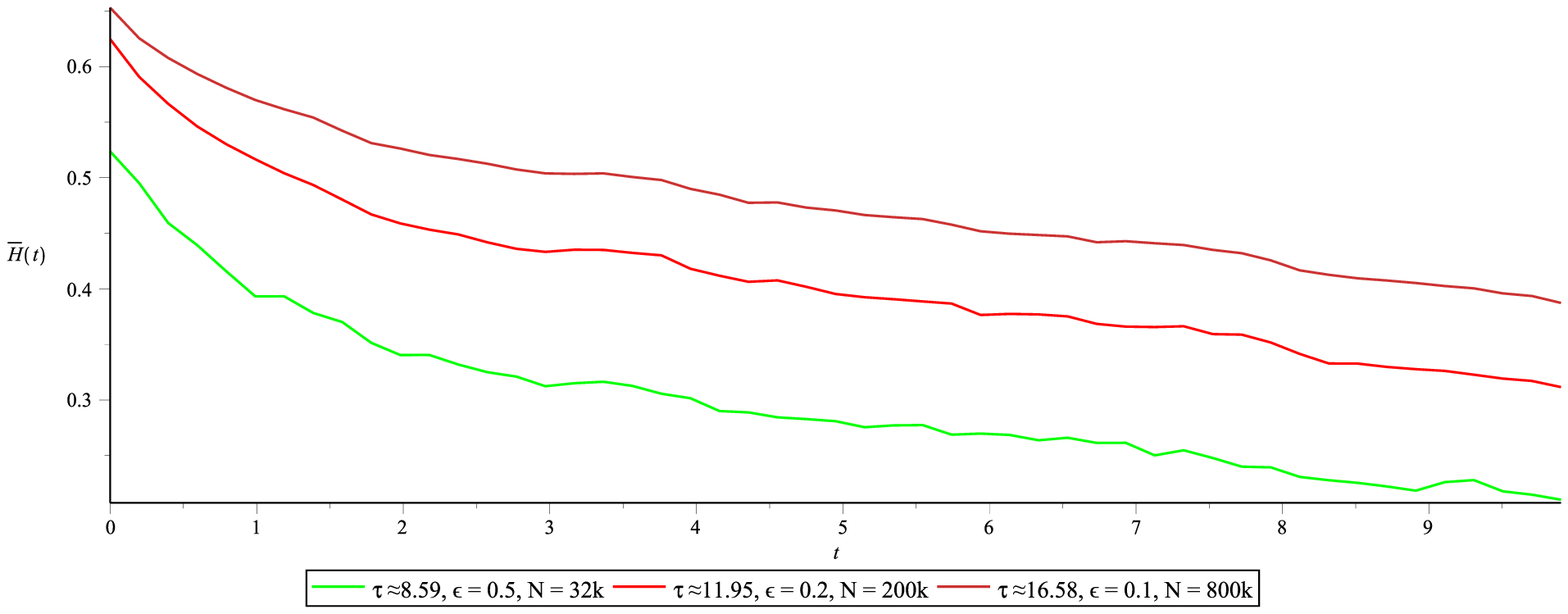}} 
\hfill
\subfloat{\includegraphics[width=0.8\linewidth, height=0.2\textheight]{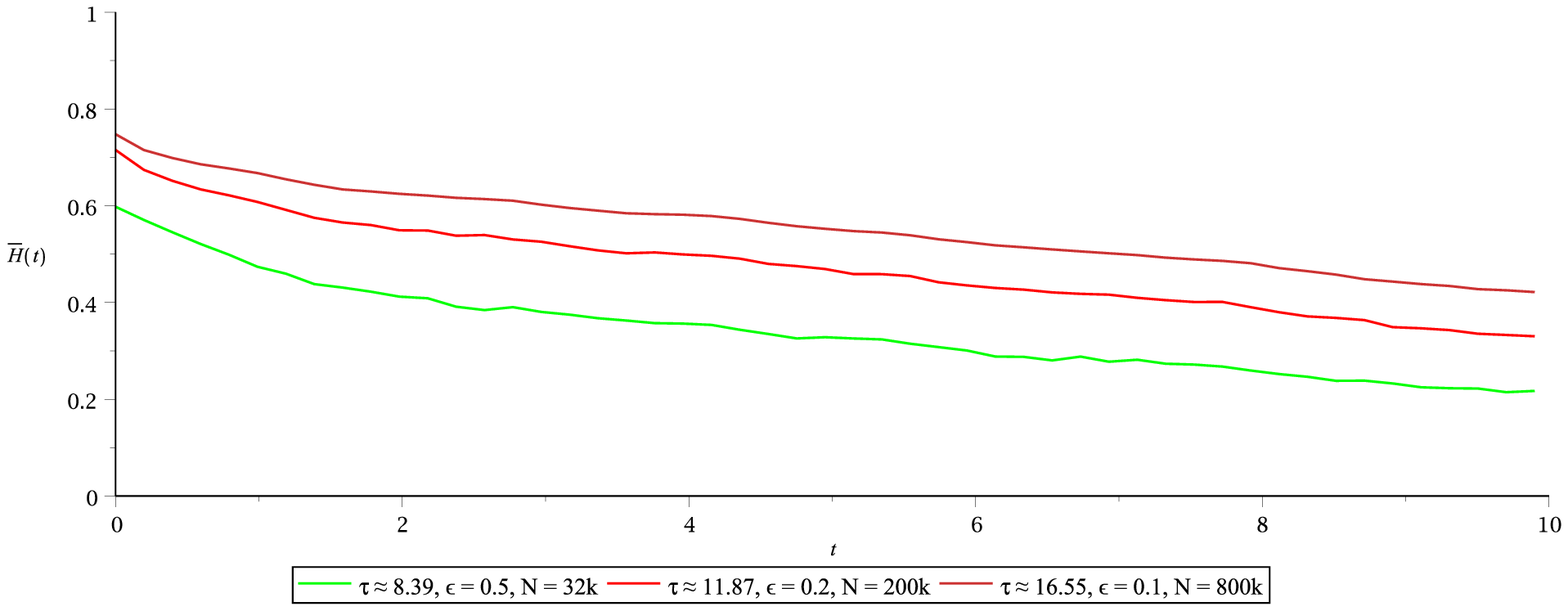}} 
\hfill
  \caption{Averages of $\bar{H}(t)$ over ten sets of phases/initial states for different coarse-graining lengths and number of test particles. From top to botom the simulations where done with increasing number of modes $M = 9, 12, 15$.
  }
  \label{diffcgl}
\end{figure}

\begin{figure}[!ht]
  \begin{minipage}{\linewidth}
  %\hfill
  \includegraphics[width=.3\linewidth]{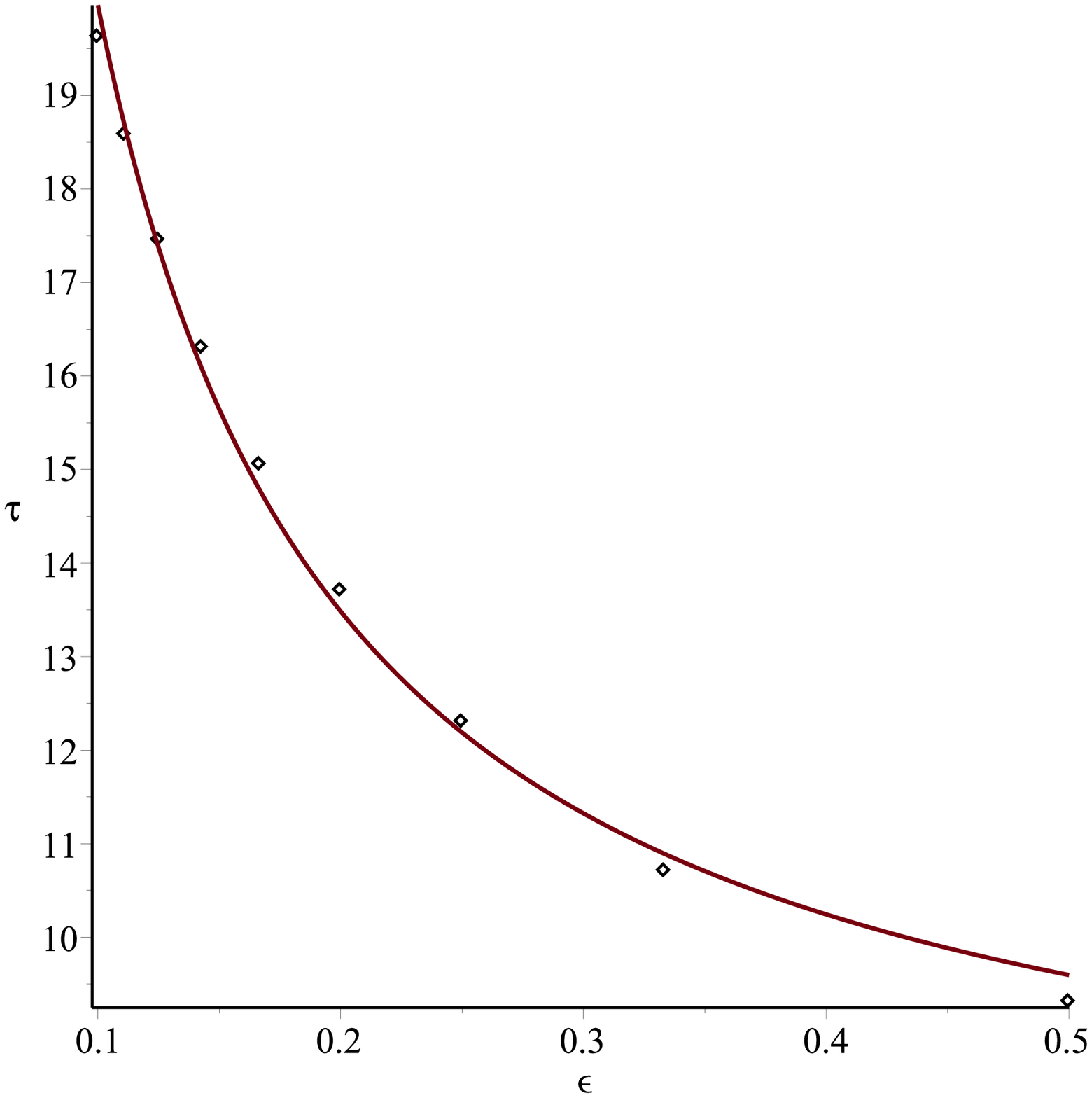}
  %\hfill
  \includegraphics[width=.3\linewidth]{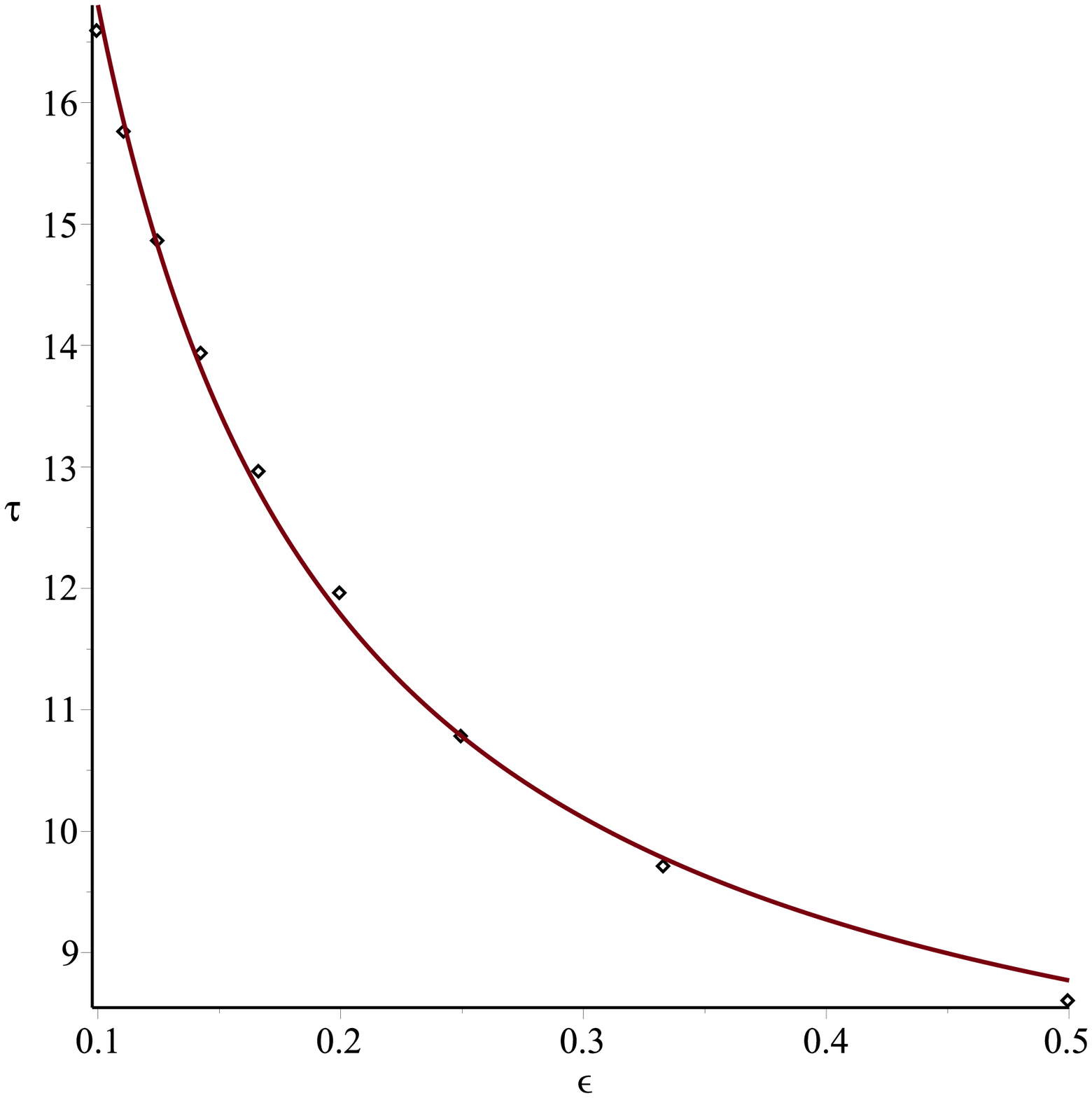}
  %\hfill
  \includegraphics[width=.3\linewidth]{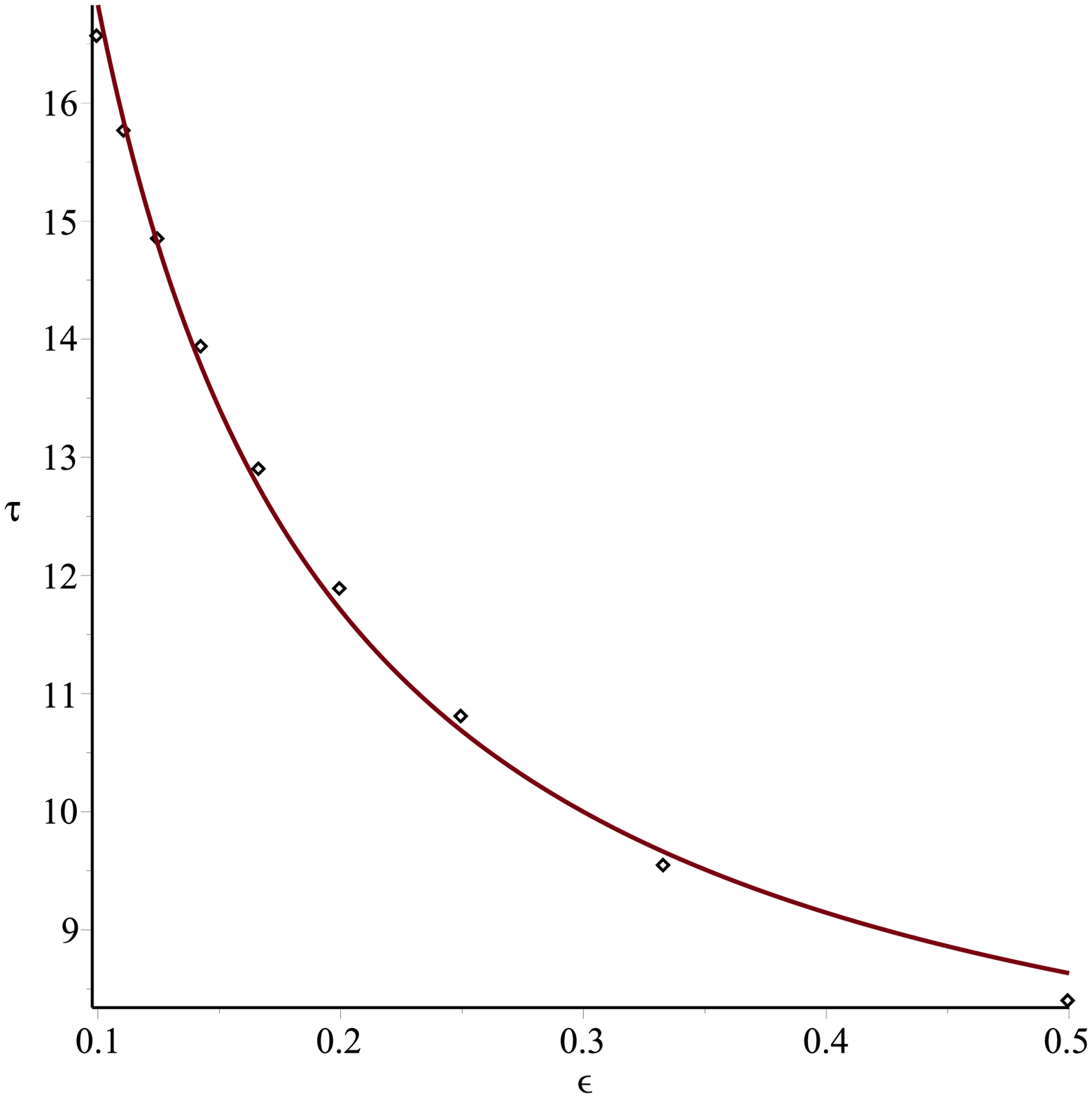}
  %\hfill
%  \includegraphics[width=.3\linewidth]{example-image-a}%
  \end{minipage}%
   \caption{$\tau$
   as a function of the
   coarse-graining length 
   for the averaged $\bar{H}(t)$ functions of the systems with 9, 12 and 15 modes respectively.}
   \label{cglxtc}
\end{figure}

We simulated the evolution of distributions with a range of values of $\epsilon$ in order to test a possible relation of the type $\tau \propto 1/\epsilon^{\alpha}$. In Figure \ref{diffcgl} we show examples of the
evolution of
$\bar{H}(t)$ for three different numbers of modes and with diminishing coarse-graining lengths. All the cases show an increment 
in the relaxation timescale as the value of $\epsilon$ decreases. 
The relation between $\tau$ and $\epsilon$ is shown in Figure \ref{cglxtc}, where a linear best fit is also shown. The function fitted is of the form 
\begin{equation}
    \tau = \frac{a}{\epsilon} + b.
\end{equation}
Our simulations clearly show that relaxation timescales are related to the coarse-graining length in the way that was suggested by earlier works \cite{bib:Valentini2001, bib:Valentini2005}. This relation can be viewed in two ways. First, a larger ensemble of trajectories and smaller cells approximate the coarse-grained distribution function to the form of the actual distribution. So the higher values of N and the smaller the cells would give a more reliable value for $\tau$. On the other hand, with too many cells the microstructure of the distribution might play a role in the evolution of $\bar{H}(t)$ through discontinuities on the integrals of Eq. (27). We take this into account in the next subsection and make a conservative choice for the value of $\epsilon$ and $N$. 

\begin{figure}

\subfloat{
  \begin{minipage}{\linewidth}
  %\hfill
  \includegraphics[width=.6\linewidth]{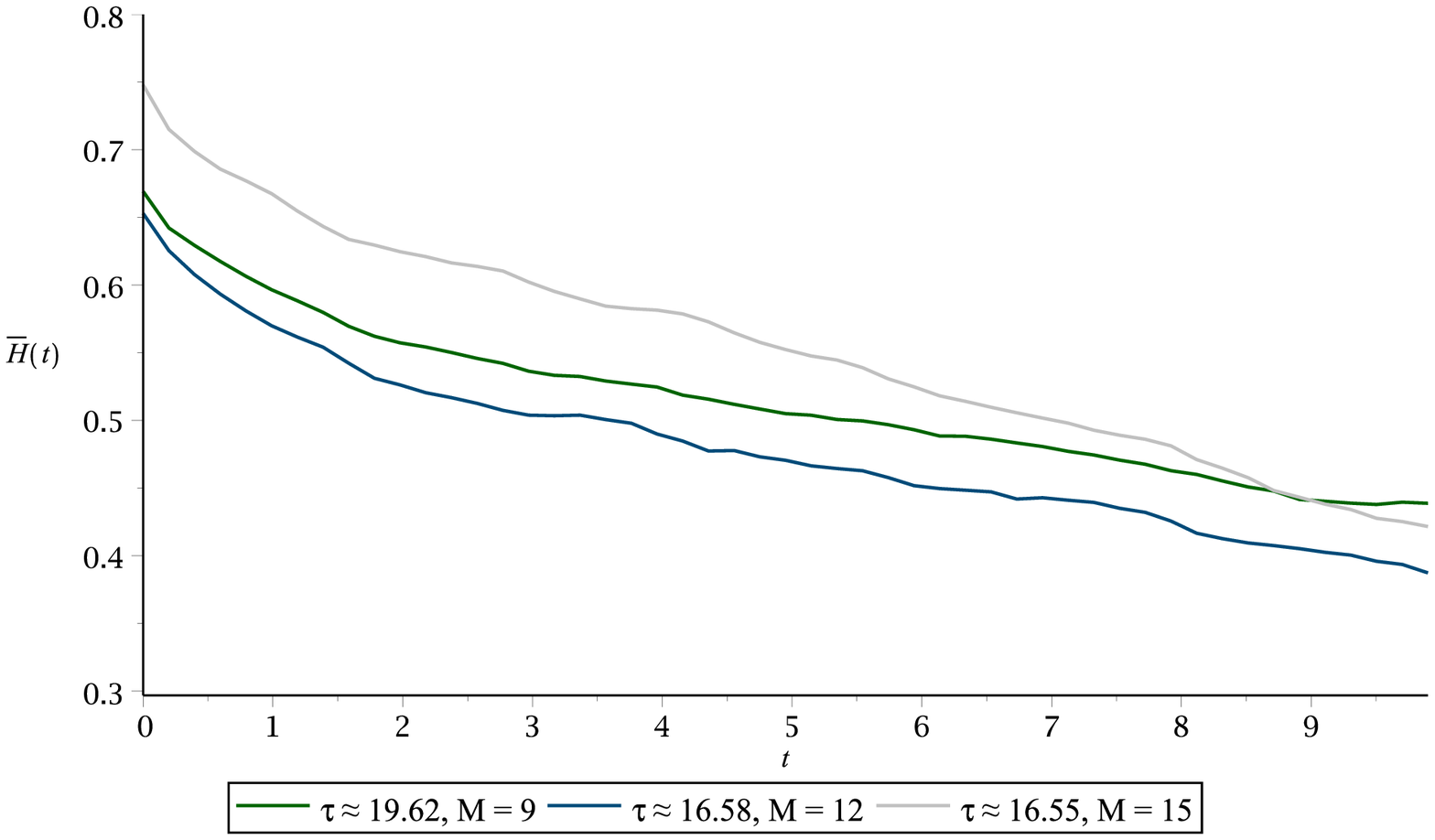}
  \includegraphics[width=.4\linewidth]{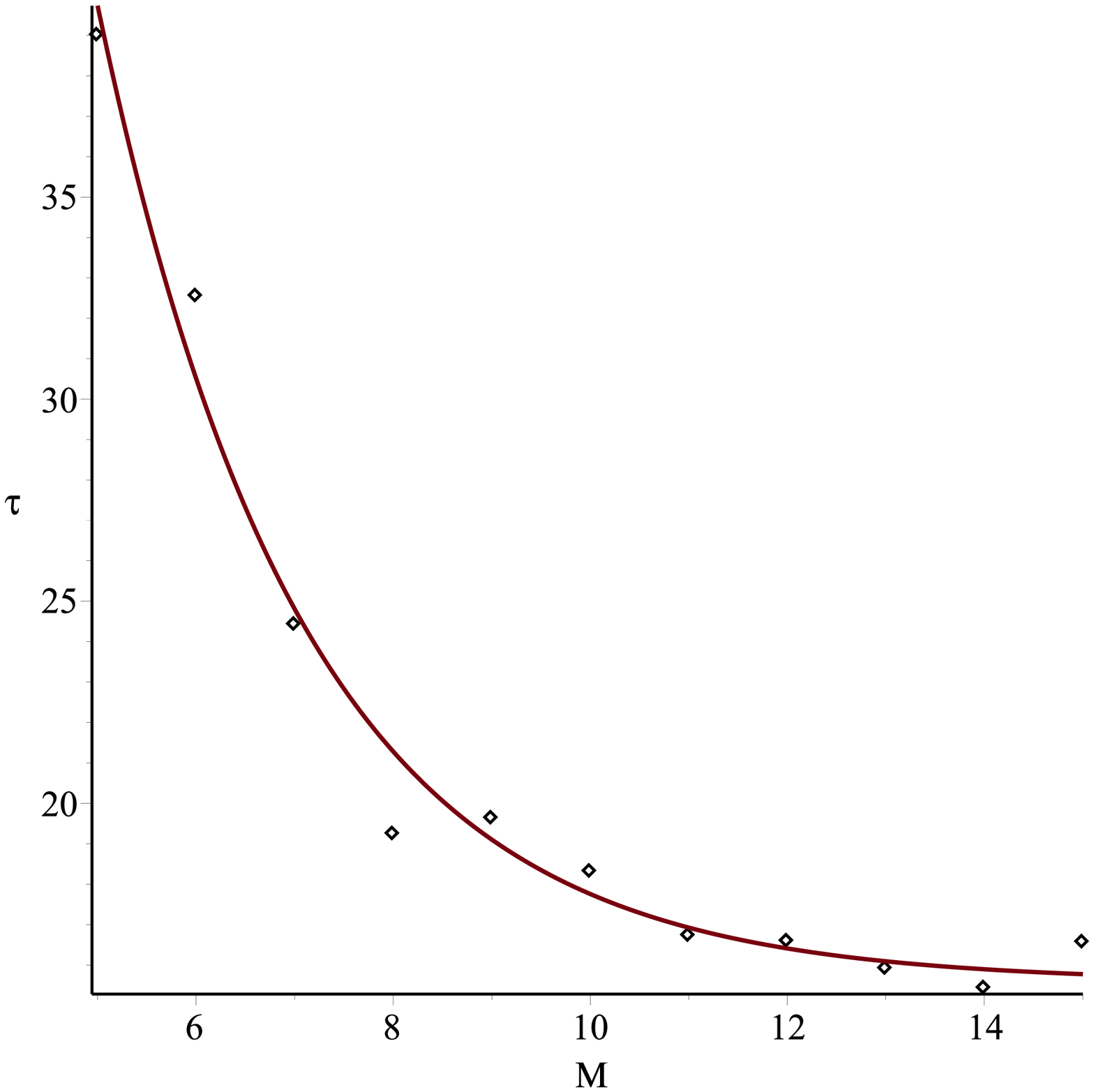}
  \end{minipage}%
} 
  \caption{Averaged $\bar{H}(t)$ functions for increasing number of modes and their estimated $\tau$ as a function of the number of modes.
  }
  \label{diffM}
\end{figure}

Another parameter whose influence has been analyzed in earlier works is the number of modes $M$. We show in Figure \ref{diffM} three examples of $\bar H(t)$ functions with different numbers of modes,
%on the wave functions
and we also show the variation of the relaxation timescale as a function of $M$. The best fit shown is of the form 
\begin{equation}
    \tau = a\exp\left(-M/b\right)+c.
\end{equation}
This simulations were done with a fixed coarse-graining length $\epsilon = 0.1$, coupling constant $\beta=0.1$, and number of test particles $N=8\times 10^5$. 

Our results are in accordance with previous works, in the sense that we also predict a decrease of the relaxation timescale with larger number of modes but gives further information because we have the possibility of calculating the $\bar{H}(t)$ for every value of $M$ ranging from 4 to 15. We then are able to provide a indication that the relation between relaxation timescales and number of modes is exponential.

\subsection{The coupling constant $\beta$}

\begin{figure}
\centering
\subfloat[$\bar{H}(t)$ for $M = 9$ modes.]{
\includegraphics[width=0.4\textwidth,height=.2\textheight]{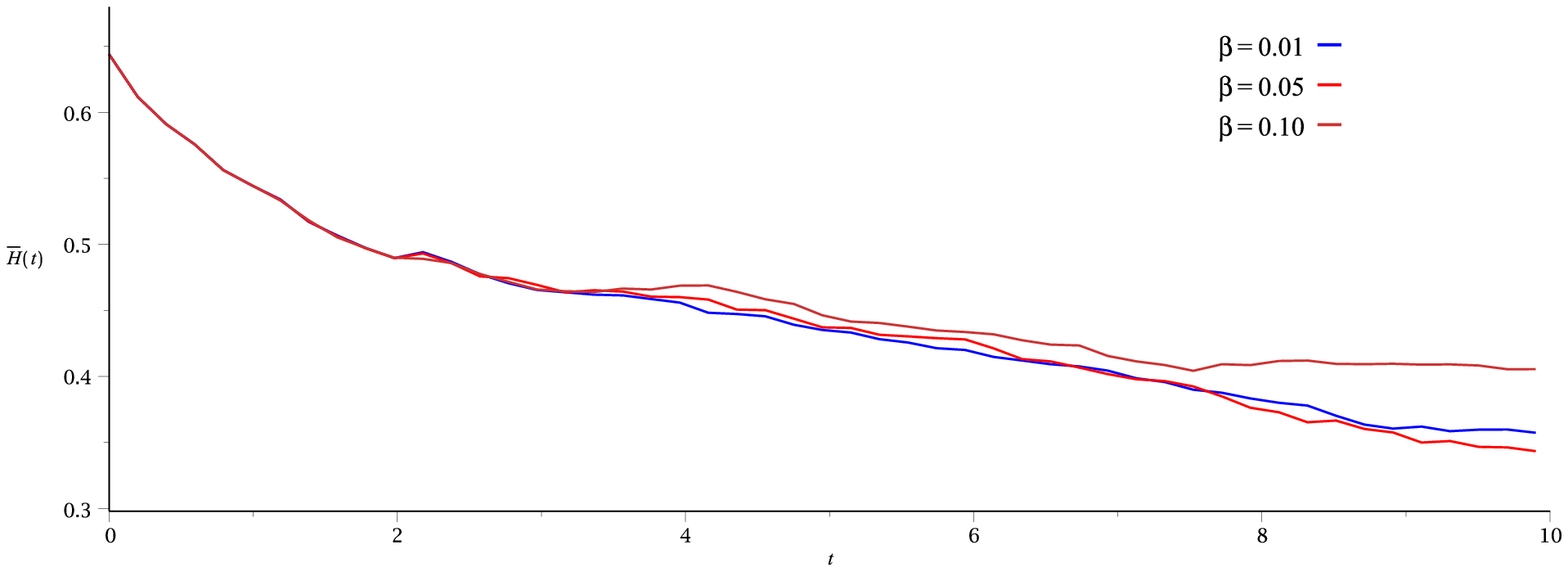}
\label{fig:subfig1}}
\qquad
\subfloat[Subfigure 2 list of figures text][$\bar{H}(t)$ for $M = 11$ modes.]{
\includegraphics[width=0.4\textwidth,height=.2\textheight]{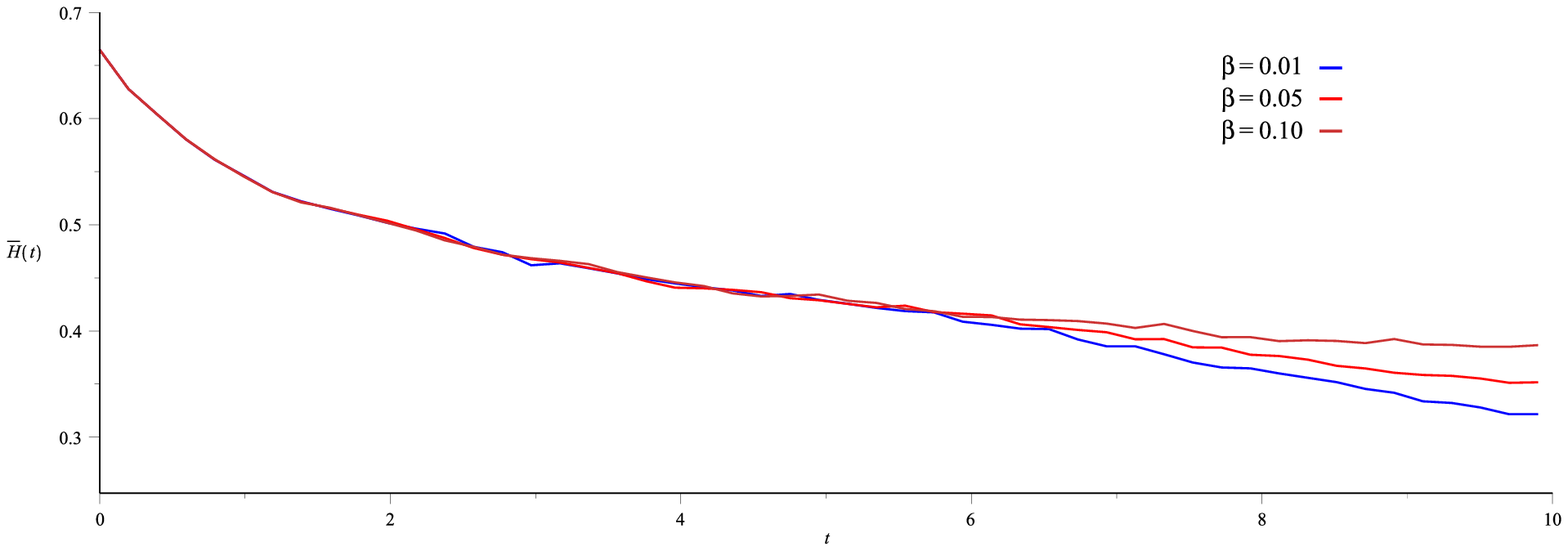}
\label{fig:subfig2}}
\hspace{0mm}
\subfloat[Subfigure 3 list of figures text][$\bar{H}(t)$ for $M = 13$ modes.]{
\includegraphics[width=0.4\textwidth,height=.2\textheight]{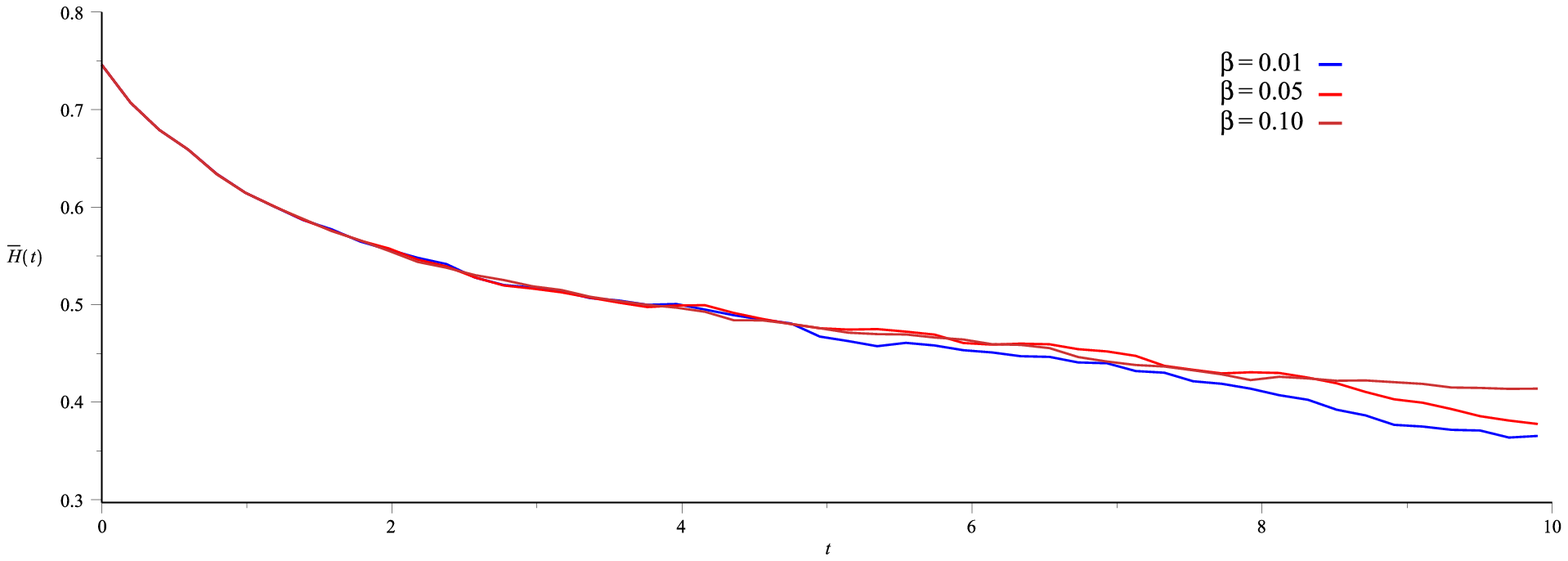}
\label{fig:subfig3}}
\qquad
\subfloat[Subfigure 4 list of figures text][$\bar{H}(t)$ for $M = 15$ modes.]{
\includegraphics[width=0.4\textwidth
,height=.2\textheight
]{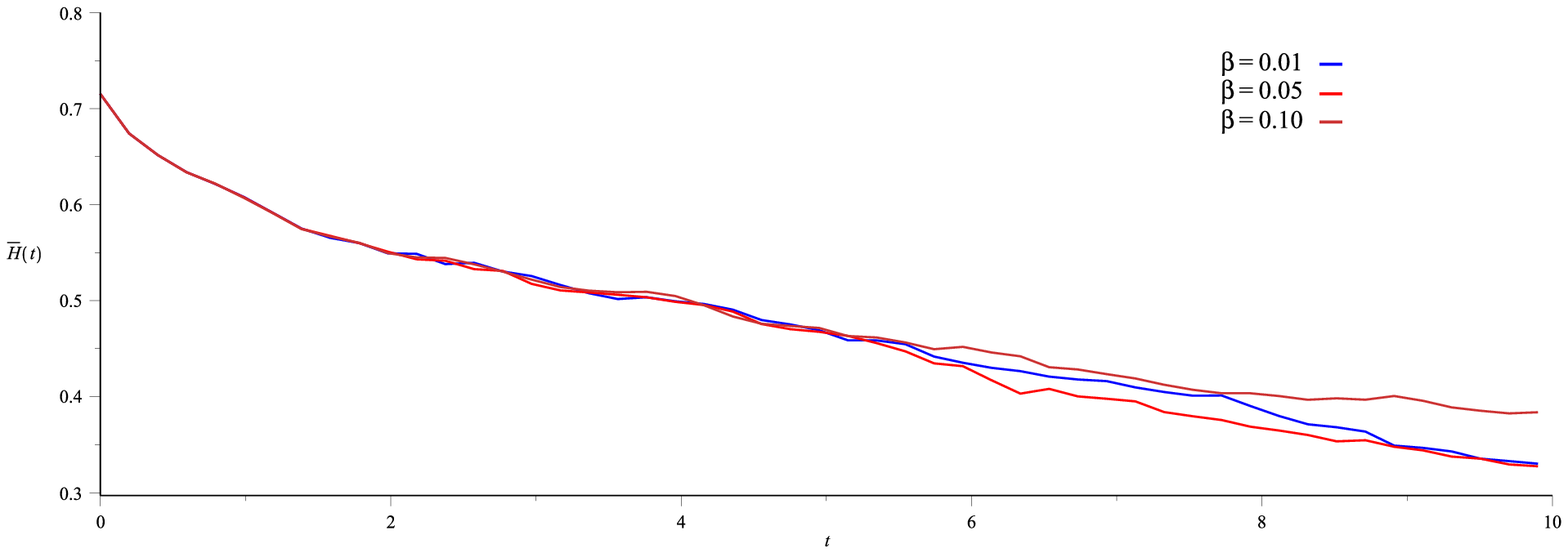}
\label{fig:subfig4}}
\caption{Each plot shows three averaged $\bar{H}(t)$ for different values of the coupling constant $\beta$.}
\label{Hbeta}
\end{figure}

It has been suggested that interactions should lead to a shorter relaxation time. To check this assertion, 
the plots of the averages of $\bar{H}(t)$ over ten sets of phases and initial states for different values of $\beta$, with $\epsilon= 0.2$ and $N= 2\times 10^5$ are presented in Figure \ref{Hbeta}. The functions decay approximately exponentially with slightly different decay rates. For all the functions checked the highest value of $\beta$ corresponds to the maximum value for the relaxation timescale $\tau$.

The best-fits for $\tau$ as a function of $\beta$ (Figure \ref{tcxbeta}) are highly nonlinear, thus showing that the role of interactions in the approach to quantum equilibrium is not as direct as previously thought. The plots suggest that, after reaching a minimum, $\tau$ grows almost monotonically with  $\beta$. This would indicate that stronger interactions, instead of accelerating relaxation, actually can delay it. 

This is our main result: interactions may not have the effect of accelerating relaxation. Further study must be done to analyze whether stronger interactions can delay relaxation providing a way in which nonequilibrium might survive some processes that were previously seen to erase it. Our plots show that the variation of the number of modes induces different variations on the data of $\tau(\beta)$ which can have both physical and numerical reasons. The fitted functions plotted represent our attempt to find a common function that best fitted all the data. The functions plotted are all polynomials of degree 5 of the form
\begin{equation}
a_0+a_1\beta+a_2\beta^2+a_3\beta^3+a_4\beta^4+a_5\beta^5,
\end{equation}
which might represent the Taylor expansion of a more complex oscillating function, 
This fits are not completely reliable as an indication of what the exact relation between $\tau$ and $\beta$ might be, but represent the first indication that the timescale can increase when the interaction is stronger. Hence, our results show a tendency for the behaviour of this system in particular that is contrary to what is generally expected. \clearpage
Further work in similar systems with more parameters are currently under way.

\subsection{Precision, Equivariance and Confidence intervals}

The results shown above were obtained by a series of steps, each with its own particular numerical limitations. First, we defined the desired the precision for the Runge-Kutta method that solved the guidance equations, which ranged from $10^{-5}$ to $10^{-15}$. 
Afterwards, in order to calculate $\bar{\rho}$ and $\bar{\psi}$ we had to define the coarse-graining length which determines the precision of the coarse-grained function with relation to the original one. In section 4.3 we showed the effect of the variation of the coarse-graining length but we took the precaution of maintaining the ratio between the number of cells and the number of calculated trajectories at $0.0125$, which is the precision for the values of the coarse-grained function. Since that is not a high precision, we tested some sample distributions of trajectories for the equivariance of the distribution with the wave-function. By equivariance, we mean the property that a distribution initially at equilibrium ($\rho = |\psi|^2$) will remain in equilibrium. The equivariance condition is mantained by our code which gives further confidence in our results.

Finally, we have the fits that give us the relaxation timescales $\tau$ and their dependence on the parameters $M, \epsilon$ and $\beta$. All the fits were made using Maple\textsuperscript{TM} and to evaluate the precision of the fitted functions we analyzed the confidence intervals on the parameters of those functions.

The fitted $\bar{H}(t)$ functions have confidence intervals for the parameter $\tau$ that are considerably high for low values of $M$, ranging from $10$ to $20\%$ of the calculated value for the relaxation timescale with the number of modes $M=4..9$. For $M=10..14$ the confidence intervals fall to the $5-15\%$ range and are of the order of $5\%$ for $M=15$. 

Those ranges indicate that the results for higher $M$ are more reliable, which could be expected if we assumed that for lower number of modes the decay to equilibrium might not be complete as indicated in earlier works \cite{bib:Abraham2014}. We did not consider the possibility of residues in the fitted $\bar{H}(t)$ function because our analysis only allowed us to observe the evolution of the trajectories inside the interval $[0, \omega^2/\beta]$. Since the highest  value of $\beta$ we included in our simulations was $0.10$ we chose to restrict the whole study to the interval $[0,10]$ (with $\omega = 1$), which is a short interval to observe either complete relaxation or the presence of residue. Our results serve to indicate the behaviour of the relaxation but further
%investigations must be done 
work is needed
to confirm them. 

\begin{figure}
\centering
\subfloat[][$M = 9$]{
  \includegraphics[width=.4\linewidth,height=.2\textheight]{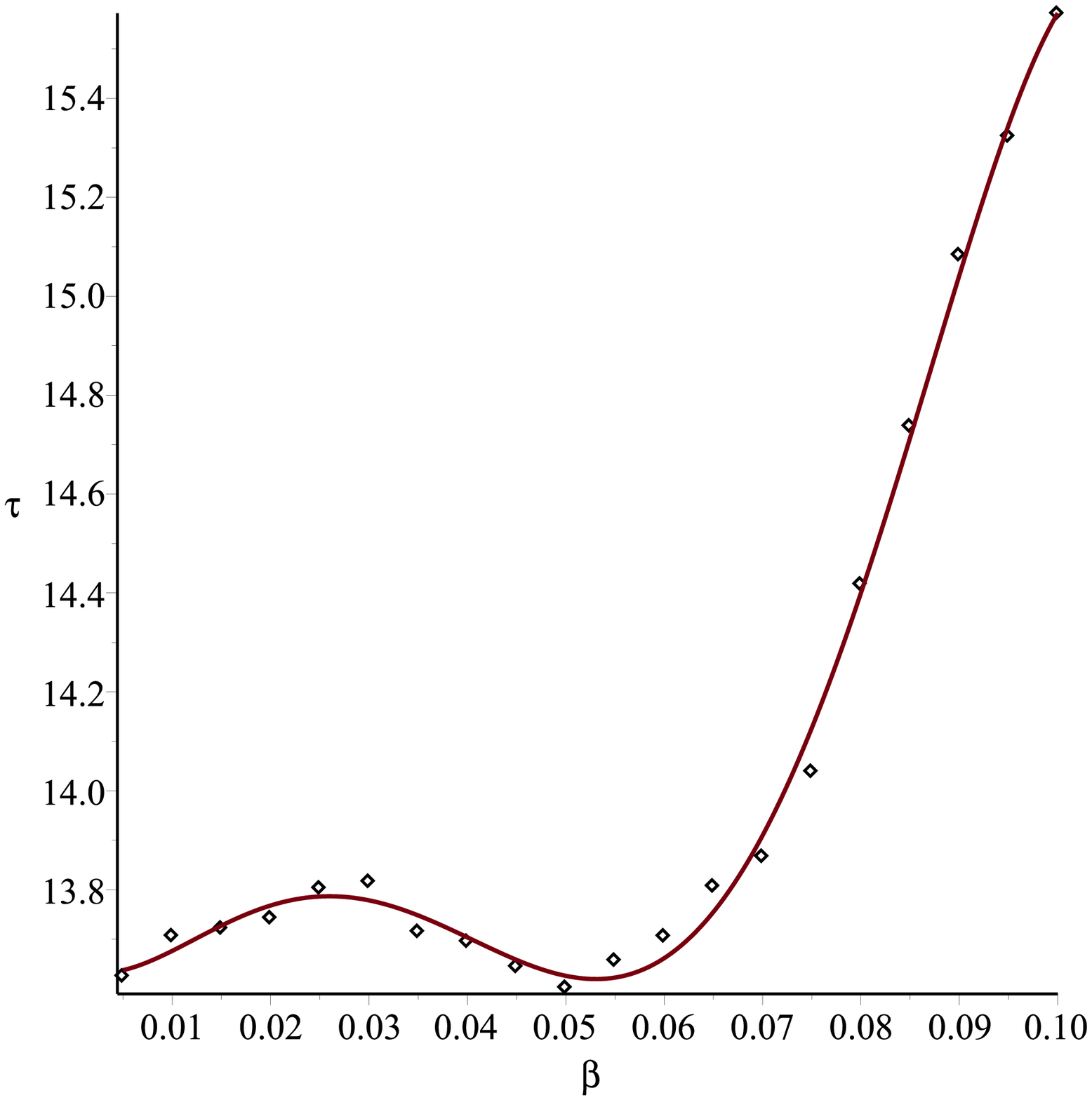}
}
\subfloat[][$M = 11$]{
  \includegraphics[width=.4\linewidth, height=.2\textheight]{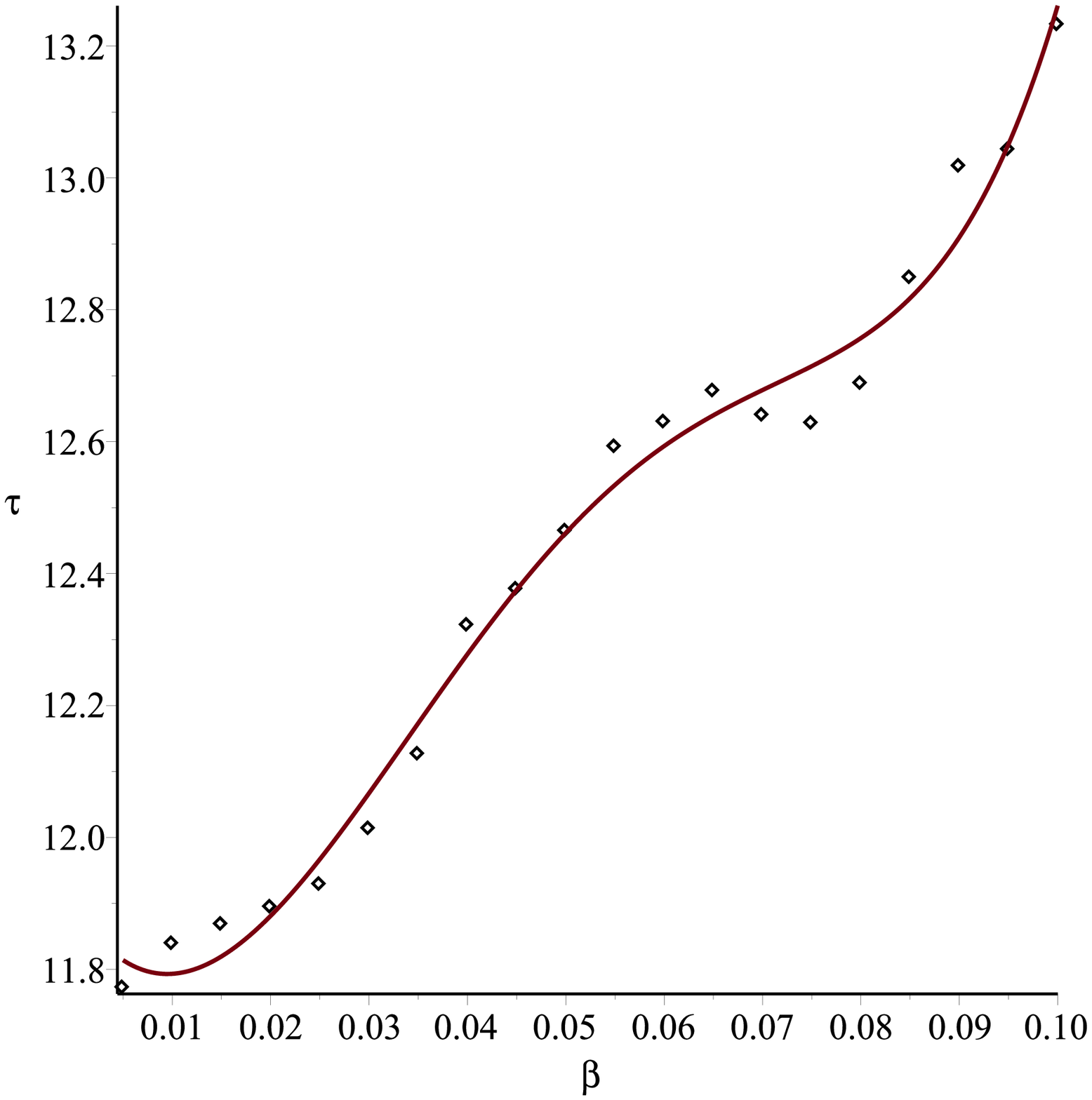}
}
\hspace{0mm}
\subfloat[][$M = 13$]{
  \includegraphics[width=.4\linewidth, height=.2\textheight]{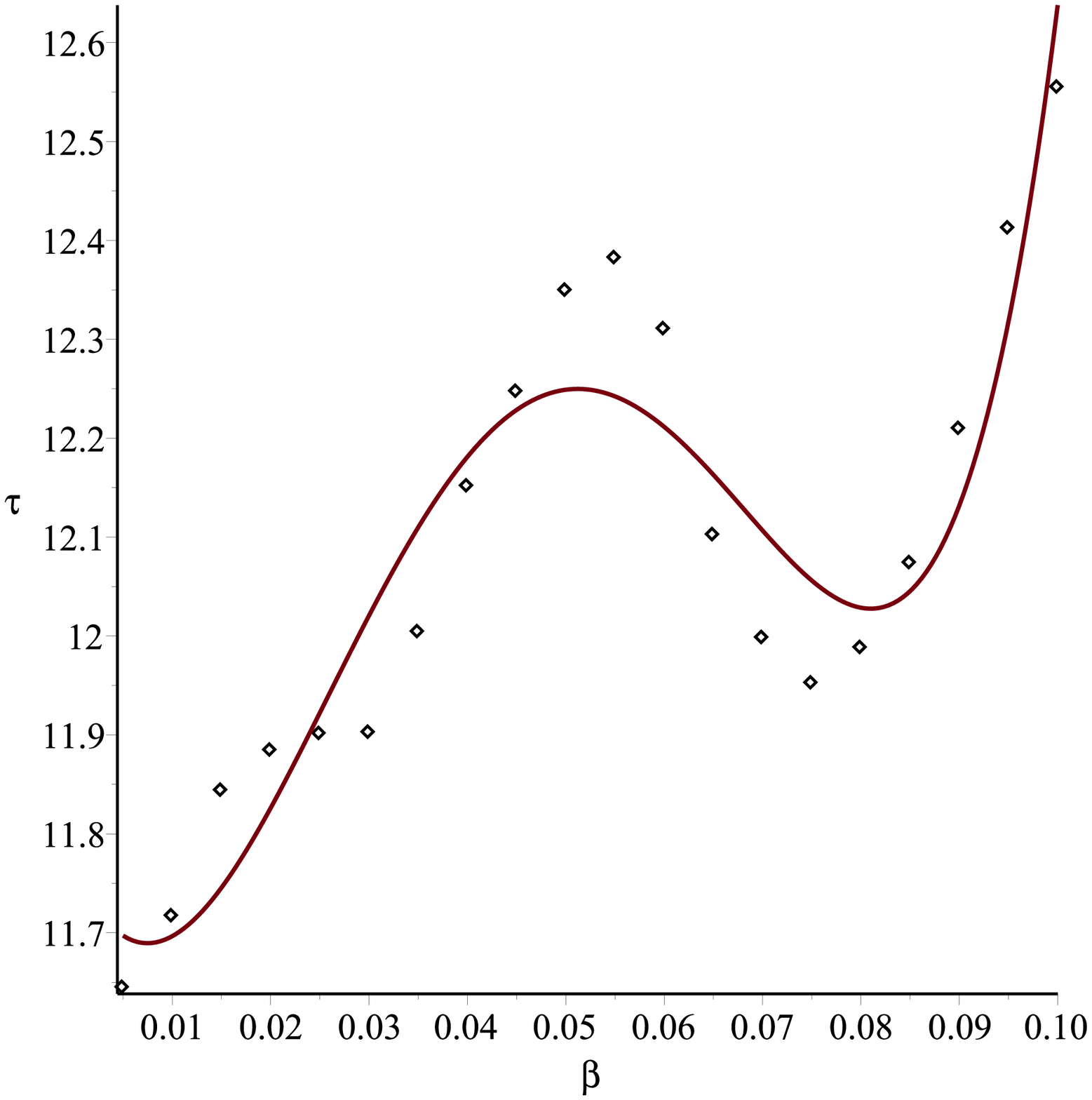}
}
\subfloat[][$M = 15$]{
  \includegraphics[width=.4\linewidth, height=.2\textheight]{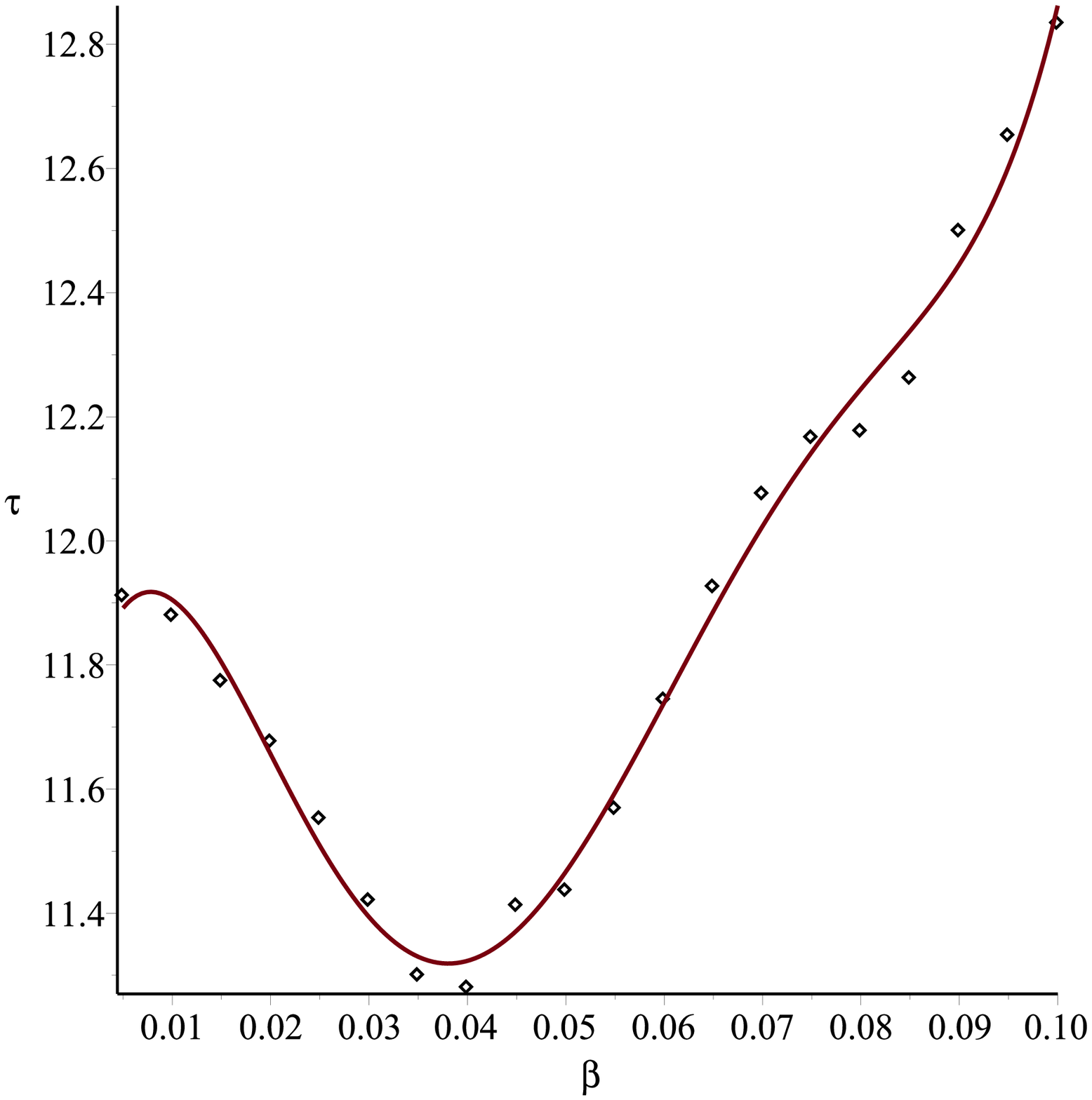}
}
\caption{$\tau$ as a function of the coupling constant $\beta$  for systems with different number of modes.}
\label{tcxbeta}
\end{figure}

Finally, the fitted functions $\tau(\epsilon)$ (Eq. (32)) and $\tau(M)$ (Eq. (33)) have confidence intervals on the parameters of the order of $10\%$ of their values, making them considerably reliable. The relation between $\tau$ and $\beta$, however, is very approximate. The polynomial form of Eq.( 34) may indicate that this is a Taylor expansion of some more complex function with fewer parameters. The plots in Figure \ref{tcxbeta} 
%serve to indicate 
suggest
that higher
\clearpage 
values of the coupling constant might be linked to slower relaxation, which is the first step to understand the role interactions may play on the decay to quantum equilibrium.

\section{Conclusions}
\label{conc}

We have studied for the first time the influence of interactions on the timescale of the relaxation process to quantum equilibrium. We have found that the timescale consistently increases with the interaction constant coupling. The increase is apparently not linear but we observed an approximately similar behaviour for a wide range of number of modes in the wave function of the system. 

Our approach has new and interesting features, including randomly generated phases and quantum numbers. This allowed us to include in the calculations of the averaged $\bar{H}(t)$ functions a variety of initial wave functions. Furthermore, we also allowed for more numbers of modes ($12$ in total) to be analyzed by not restraining %constraining 
to symmetric superpositions. This
permited 
% meant that we could 
to 
check in detail
%with a larger number of parameters 
the dependence of $\tau$ with $M$, indicating the possibility of a exponential relation. Varying the coarse-graining length we also found a relation between $\tau$ and $\epsilon$ of the form $\tau=a/\epsilon + b$, in line 
%similar 
to the one 
%what was 
predicted in earlier works \cite{bib:Valentini2001, bib:Valentini2005}.

The numerical analysis covered a wide range of parameter values and in all the simulations the approximate exponential decay of the $\bar{H}(t)$ was observed with the predicted form $H_0\exp{(-t/\tau)}$. This confirms the indications in all the previous simulations but now for a system with explicit time-dependent 
interaction.

The system studied here was %intended to be the simplest 
%possible introduction 
chosen because it allows 
to introduce a
time-dependent interaction in the study of the relaxation process in a simple way. It has been   claimed that strong interactions might be responsible for `washing out' the nonequilibrium through the evolution of the early Universe. Although this might be true for most of the systems that we observe today (since no current experiment has observed violations of the Born rule), our results indicate the possibility that some form of interactions might delay and even prevent complete relaxation.
\clearpage
Although our analysis has  (mainly numerical and computational)
limitations, it is still a first step in the direction of actually describing a mechanism through which interactions may contribute to affect relaxation of some relic systems in the the way proposed in \cite{bib:Underwood2015}. Such systems could have avoided complete relaxation in the very early Universe and could have nontrivial effects on the CMB or on other relic systems \cite{bib:Valentini2001, bib:Valentini2007, bib:Colin2015, bib:Colin2016, bib:Vitenti2019}.

The numerical and theoretical analysis done so far in the literature have been focused on some important assumptions about the cosmology of the early Universe. First, only the evolution of the degrees of freedom of one massless free scalar field  on an expanding, not inflationary, background \cite{bib:Colin2015, bib:Colin2016}. The scenario is based on the evolution of a nonequilibrium distribution in a pre-inflationary era and in a \textit{freezing inequality} \cite{bib:Valentini2010} that indicates that some modes that exit the Hubble radius before or right at the beginning of inflation might carry a nonequilibrium signature that could be detected in the CMB \cite{bib:Vitenti2019} in relic particles that decoupled very early after inflation (the gravitino being the most likely candidate \cite{bib:Underwood2015}). 

There are multiple reasons to question if the above scenario is feasible: the uncertainty about the physics in a pre-inflationary era, the dependence of the freezing of super Hubble modes on the assumption that they are initially in the Bunch-Davies vacuum, the difficulties associated to actually predicting what would be the exact effects of nonequilibrium on the CMB and how to separate those effects from the ones predicted by different models of early Universe cosmology, among others. A different approach would be to consider the application of the de Broglie-Bohm theory to the whole cosmological model \cite{bib:Pinto-Neto2005, bib:Peter2008, bib:Pinto-Neto2012, bib:Pinto-Neto2013, bib:Pinto-Neto2016, bib:Pinto-Neto2018}, which gives rise to a number of possible cosmologies currently under study. 

That approach has been developed initially as a way to substitute the initial singularity of the standard $\Lambda$CDM model with a bounce that occurs due to quantum effects that are only present in the pilot-wave picture but in all the studies done so far the initial quantum state of the Universe were always considered to be at equilibrium. A number of questions can be made on the possible effects that a nonequilibrium initial distribution could have on the power-spectrum predicted by bouncing models and on the particle produced in these scenarios. 

Another possibility would be to consider the evolution of nonequilibrium for initial states that differ from the Bunch-Davies vacuum \cite{bib:Jerome2000, bib:Greene2006}. Using the pilot-wave approach we could derive different forms for the freezing inequality for different initial states and provide a way through which we could test the validity of nonvacuum and nonequilibrium initial states. 

These questions will be subject to future works where we can apply the numerical analysis presented here to different cosmological models and further constrain the possible ways through which violations of the Born rule could be tested today. 

\section*{Acknowledgment}

We would like to thank CNPq and CAPES of Brazil for financial support. F. B. Lustosa would also like to thank A. Valentini, N. G. Underwood and A. Kandhadai for helpful discussions.S. Colin thanks FQXi for its financial support through the mini-grants FQXi-MGA-1804 and FQXi-MGA-1915, Jurgen Theiss for administering these grants,
and the hospitality of the CBPF during two research visits (in particular, that of Nelson Pinto-Neto).

\section*{References}
\bibliography{mainexemplos} 

\begin{thebibliography}{10}
\expandafter\ifx\csname url\endcsname\relax
  \def\url#1{\texttt{#1}}\fi
\expandafter\ifx\csname urlprefix\endcsname\relax\def\urlprefix{URL }\fi
\expandafter\ifx\csname href\endcsname\relax
  \def\href#1#2{#2} \def\path#1{#1}\fi

\bibitem{bib:Bohm1951a}
D.~Bohm, \href{https://link.aps.org/doi/10.1103/PhysRev.85.166}{A suggested
  interpretation of the quantum theory in terms of "hidden" variables. i},
  Phys. Rev. 85 (1952) 166--179.
\newblock \href {http://dx.doi.org/10.1103/PhysRev.85.166}
  {\path{doi:10.1103/PhysRev.85.166}}.
\newline\urlprefix\url{https://link.aps.org/doi/10.1103/PhysRev.85.166}

\bibitem{bib:Bohm1951b}
D.~Bohm, A suggested interpretation of the quantum theory in terms of hidden
  variables. 2., Phys. Rev. 85 (1952) 180--193.
\newblock \href {http://dx.doi.org/10.1103/PhysRev.85.180}
  {\path{doi:10.1103/PhysRev.85.180}}.

\bibitem{bib:Bohm1993}
D.~Bohm, B.~J. Hiley, The Uundivided Universe: An Ontological Intepretation of
  Quantum Theory, Routledge and Kegan Paul, London, 1993.

\bibitem{bib:Holland1995}
P.~Holland, \href{https://books.google.com.br/books?id=BsEfVBzToRMC}{The
  Quantum Theory of Motion: An Account of the de Broglie-Bohm Causal
  Interpretation of Quantum Mechanics}, Cambridge University Press, 1995.
\newline\urlprefix\url{https://books.google.com.br/books?id=BsEfVBzToRMC}

\bibitem{bib:Valentini1990}
A.~Valentini, {Signal-locality, uncertainty, and the subquantum H-theorem. I},
  Phys. Lett. A156 (1991) 5--11.
\newblock \href {http://dx.doi.org/10.1016/0375-9601(91)90116-P}
  {\path{doi:10.1016/0375-9601(91)90116-P}}.

\bibitem{bib:Valentini1991}
A.~Valentini, {Signal-locality, uncertainty, and the subquantum H-theorem. II},
  Phys. Lett. A158 (1991) 1--8.
\newblock \href {http://dx.doi.org/10.1016/0375-9601(91)90330-B}
  {\path{doi:10.1016/0375-9601(91)90330-B}}.

\bibitem{bib:Valentini1992}
A.~Valentini, \href{http://hdl.handle.net/20.500.11767/4334}{{On the pilot-wave
  theory of classical, quantum and subquantum physics}}, Ph.D. thesis, SISSA,
  Trieste (1992).
\newline\urlprefix\url{http://hdl.handle.net/20.500.11767/4334}

\bibitem{bib:Valentini2001}
A.~{Valentini}, {Hidden Variables, Statistical Mechanics and the Early
  Universe}, Vol. 574, 2001, p. 165.

\bibitem{bib:Valentini2005}
A.~Valentini, H.~Westman, {Dynamical origin of quantum probabilities}, Proc.
  Roy. Soc. Lond. A461 (2005) 253--272.
\newblock \href {http://arxiv.org/abs/quant-ph/0403034}
  {\path{arXiv:quant-ph/0403034}}, \href
  {http://dx.doi.org/10.1098/rspa.2004.1394}
  {\path{doi:10.1098/rspa.2004.1394}}.

\bibitem{bib:Towler2012}
M.~D. {Towler}, N.~J. {Russell}, A.~{Valentini}, {Time scales for dynamical
  relaxation to the Born rule}, Proceedings of the Royal Society of London
  Series A 468 (2012) 990--1013.
\newblock \href {http://arxiv.org/abs/1103.1589} {\path{arXiv:1103.1589}},
  \href {http://dx.doi.org/10.1098/rspa.2011.0598}
  {\path{doi:10.1098/rspa.2011.0598}}.

\bibitem{bib:Valentini2019}
A.~Valentini, {Foundations of statistical mechanics and the status of the Born
  rule in de Broglie-Bohm pilot-wave theory}, 2020, pp. 423--477.
\newblock \href {http://arxiv.org/abs/1906.10761} {\path{arXiv:1906.10761}},
  \href {http://dx.doi.org/10.1142/9789811211720\_0012}
  {\path{doi:10.1142/9789811211720\_0012}}.

\bibitem{bib:Valentini1996}
A.~Valentini, \href{https://doi.org/10.1007/978-94-015-8715-0_3}{Pilot-Wave
  Theory of Fields, Gravitation and Cosmology}, Springer Netherlands,
  Dordrecht, 1996, pp. 45--66.
\newblock \href {http://dx.doi.org/10.1007/978-94-015-8715-0_3}
  {\path{doi:10.1007/978-94-015-8715-0_3}}.
\newline\urlprefix\url{https://doi.org/10.1007/978-94-015-8715-0_3}

\bibitem{bib:Valentini2001uq}
A.~Valentini, {Signal locality and subquantum information in deterministic
  hidden variables theories}, in: {NATO Advanced Research Workshop on Modality,
  Probability, and Bell's Theorems}, 2001.
\newblock \href {http://arxiv.org/abs/quant-ph/0112151}
  {\path{arXiv:quant-ph/0112151}}.

\bibitem{bib:Valentini2004ep}
A.~Valentini, {Black holes, information loss, and hidden variables}\href
  {http://arxiv.org/abs/hep-th/0407032} {\path{arXiv:hep-th/0407032}}.

\bibitem{bib:Valentini2004xv}
A.~Valentini, {Extreme test of quantum theory with black holes}\href
  {http://arxiv.org/abs/astro-ph/0412503} {\path{arXiv:astro-ph/0412503}}.

\bibitem{bib:Valentini2005st}
A.~Valentini, {Hidden variables and the large-scale structure of
  spacetime}\href {http://arxiv.org/abs/quant-ph/0504011}
  {\path{arXiv:quant-ph/0504011}}.

\bibitem{bib:Valentini2007}
A.~Valentini, {Astrophysical and cosmological tests of quantum theory}, J.
  Phys. A40 (2007) 3285--3303.
\newblock \href {http://arxiv.org/abs/hep-th/0610032}
  {\path{arXiv:hep-th/0610032}}, \href
  {http://dx.doi.org/10.1088/1751-8113/40/12/S24}
  {\path{doi:10.1088/1751-8113/40/12/S24}}.

\bibitem{bib:Valentini2008}
A.~Valentini, {De Broglie-Bohm Prediction of Quantum Violations for
  Cosmological Super-Hubble Modes}\href {http://arxiv.org/abs/0804.4656}
  {\path{arXiv:0804.4656}}.

\bibitem{bib:Valentini2010}
A.~Valentini, {Inflationary Cosmology as a Probe of Primordial Quantum
  Mechanics}, Phys. Rev. D82 (2010) 063513.
\newblock \href {http://arxiv.org/abs/0805.0163} {\path{arXiv:0805.0163}},
  \href {http://dx.doi.org/10.1103/PhysRevD.82.063513}
  {\path{doi:10.1103/PhysRevD.82.063513}}.

\bibitem{bib:Colin2013}
S.~Colin, A.~Valentini, {Mechanism for the suppression of quantum noise at
  large scales on expanding space}, Phys. Rev. D88 (2013) 103515.
\newblock \href {http://arxiv.org/abs/1306.1579} {\path{arXiv:1306.1579}},
  \href {http://dx.doi.org/10.1103/PhysRevD.88.103515}
  {\path{doi:10.1103/PhysRevD.88.103515}}.

\bibitem{bib:Underwood2015}
N.~G. Underwood, A.~Valentini,
  \href{http://link.aps.org/doi/10.1103/PhysRevD.92.063531}{Quantum field
  theory of relic nonequilibrium systems}, Phys. Rev. D 92 (2015) 063531.
\newblock \href {http://dx.doi.org/10.1103/PhysRevD.92.063531}
  {\path{doi:10.1103/PhysRevD.92.063531}}.
\newline\urlprefix\url{http://link.aps.org/doi/10.1103/PhysRevD.92.063531}

\bibitem{bib:Underwood2017}
N.~G. Underwood, A.~Valentini, {Anomalous spectral lines and relic quantum
  nonequilibrium}\href {http://arxiv.org/abs/1609.04576}
  {\path{arXiv:1609.04576}}.

\bibitem{bib:Kandhadai2016}
A.~Kandhadai, A.~Valentini, {Perturbations and quantum relaxation}, Found.
  Phys. 49~(1) (2019) 1--23.
\newblock \href {http://arxiv.org/abs/1609.04485} {\path{arXiv:1609.04485}},
  \href {http://dx.doi.org/10.1007/s10701-018-0227-3}
  {\path{doi:10.1007/s10701-018-0227-3}}.

\bibitem{bib:Colin2011}
S.~Colin, Relaxation to quantum equilibrium for dirac fermions in the de
  broglie-bohm pilot-wave theory, Proc. Roy. Soc. Lond. A468 (2012) 1116.
\newblock \href {http://arxiv.org/abs/1108.5496} {\path{arXiv:1108.5496}},
  \href {http://dx.doi.org/10.1098/rspa.2011.0549}
  {\path{doi:10.1098/rspa.2011.0549}}.

\bibitem{bib:Abraham2014}
E.~Abraham, S.~Colin, A.~Valentini,
  \href{http://stacks.iop.org/1751-8121/47/i=39/a=395306}{Long-time relaxation
  in pilot-wave theory}, Journal of Physics A: Mathematical and Theoretical
  47~(39) (2014) 395306.
\newline\urlprefix\url{http://stacks.iop.org/1751-8121/47/i=39/a=395306}

\bibitem{bib:Valentini2015}
A.~Valentini, {Statistical anisotropy and cosmological quantum relaxation}\href
  {http://arxiv.org/abs/1510.02523} {\path{arXiv:1510.02523}}.

\bibitem{bib:Colin2015}
S.~{Colin}, A.~{Valentini}, {Primordial quantum nonequilibrium and large-scale
  cosmic anomalies}, Physical Review D 92~(4) (2015) 043520.
\newblock \href {http://arxiv.org/abs/1407.8262} {\path{arXiv:1407.8262}},
  \href {http://dx.doi.org/10.1103/PhysRevD.92.043520}
  {\path{doi:10.1103/PhysRevD.92.043520}}.

\bibitem{bib:Colin2016}
S.~{Colin}, A.~{Valentini}, {Robust predictions for the large-scale
  cosmological power deficit from primordial quantum nonequilibrium},
  International Journal of Modern Physics D 25 (2016) 1650068.
\newblock \href {http://arxiv.org/abs/1510.03508} {\path{arXiv:1510.03508}},
  \href {http://dx.doi.org/10.1142/S0218271816500681}
  {\path{doi:10.1142/S0218271816500681}}.

\bibitem{bib:Vitenti2019}
S.~D.~P. Vitenti, P.~Peter, A.~Valentini, {Modeling the large-scale power
  deficit with smooth and discontinuous primordial spectra}, Phys. Rev.
  D100~(4) (2019) 043506.
\newblock \href {http://arxiv.org/abs/1901.08885} {\path{arXiv:1901.08885}},
  \href {http://dx.doi.org/10.1103/PhysRevD.100.043506}
  {\path{doi:10.1103/PhysRevD.100.043506}}.

\bibitem{bib:Underwood2018}
N.~G. Underwood, {Extreme quantum nonequilibrium, nodes, vorticity, drift and
  relaxation retarding states}, J. Phys. A51~(5) (2018) 055301.
\newblock \href {http://arxiv.org/abs/1705.06757} {\path{arXiv:1705.06757}},
  \href {http://dx.doi.org/10.1088/1751-8121/aa9e97}
  {\path{doi:10.1088/1751-8121/aa9e97}}.

\bibitem{bib:Valentini2014}
A.~Valentini, {Trans-Planckian fluctuations and the stability of quantum
  mechanics}\href {http://arxiv.org/abs/1409.7467} {\path{arXiv:1409.7467}}.

\bibitem{bib:Durr1992}
D.~{D{\"u}rr}, S.~{Goldstein}, N.~{Zangh{\'\i}}, {Quantum equilibrium and the
  origin of absolute uncertainty}, Journal of Statistical Physics 67~(5-6)
  (1992) 843--907.
\newblock \href {http://arxiv.org/abs/quant-ph/0308039}
  {\path{arXiv:quant-ph/0308039}}, \href {http://dx.doi.org/10.1007/BF01049004}
  {\path{doi:10.1007/BF01049004}}.

\bibitem{bib:Durr2009}
D.~D{\"u}rr, S.~Teufel,
  \href{https://books.google.com.br/books?id=UdBptQAACAAJ}{Bohmian Mechanics:
  The Physics and Mathematics of Quantum Theory}, Fundamental Theories of
  Physics, Springer, 2009.
\newline\urlprefix\url{https://books.google.com.br/books?id=UdBptQAACAAJ}

\bibitem{bib:Goldstein2017}
S.~Goldstein, Bohmian mechanics, in: E.~N. Zalta (Ed.), The Stanford
  Encyclopedia of Philosophy, summer 2017 Edition, Metaphysics Research Lab,
  Stanford University, 2017.

\bibitem{bib:Tumulka2018}
R.~Tumulka, {On Bohmian Mechanics, Particle Creation, and Relativistic
  Space-Time: Happy 100th Birthday, David Bohm!}, Entropy 20~(6) (2018) 462.
\newblock \href {http://arxiv.org/abs/1804.08853} {\path{arXiv:1804.08853}},
  \href {http://dx.doi.org/10.3390/e20060462} {\path{doi:10.3390/e20060462}}.

\bibitem{bib:Contopoulos2012}
G.~Contopoulos, N.~Delis, C.~Efthymiopoulos,
  \href{https://doi.org/10.1088%2F1751-8113%2F45%2F16%2F165301}{Order in de
  broglie{\textendash}bohm quantum mechanics}, Journal of Physics A:
  Mathematical and Theoretical 45~(16) (2012) 165301.
\newblock \href {http://dx.doi.org/10.1088/1751-8113/45/16/165301}
  {\path{doi:10.1088/1751-8113/45/16/165301}}.
\newline\urlprefix\url{https://doi.org/10.1088%2F1751-8113%2F45%2F16%2F165301}

\bibitem{bib:Frisk1997}
H.~Frisk,
  \href{http://www.sciencedirect.com/science/article/pii/S0375960197000443}{Properties
  of the trajectories in bohmian mechanics}, Physics Letters A 227~(3) (1997)
  139 -- 142.
\newblock \href
  {http://dx.doi.org/http://dx.doi.org/10.1016/S0375-9601(97)00044-3}
  {\path{doi:http://dx.doi.org/10.1016/S0375-9601(97)00044-3}}.
\newline\urlprefix\url{http://www.sciencedirect.com/science/article/pii/S0375960197000443}

\bibitem{bib:Wu1999}
H.~Wu, D.~Sprung,
  \href{http://www.sciencedirect.com/science/article/pii/S0375960199006295}{Quantum
  chaos in terms of bohm trajectories1original received date 1 august 19951},
  Physics Letters A 261~(3) (1999) 150 -- 157.
\newblock \href
  {http://dx.doi.org/https://doi.org/10.1016/S0375-9601(99)00629-5}
  {\path{doi:https://doi.org/10.1016/S0375-9601(99)00629-5}}.
\newline\urlprefix\url{http://www.sciencedirect.com/science/article/pii/S0375960199006295}

\bibitem{bib:Wisniacki2005}
{Wisniacki, D. A.}, {Pujals, E. R.},
  \href{https://doi.org/10.1209/epl/i2005-10085-3}{Motion of vortices implies
  chaos in bohmian mechanics}, Europhys. Lett. 71~(2) (2005) 159--165.
\newblock \href {http://dx.doi.org/10.1209/epl/i2005-10085-3}
  {\path{doi:10.1209/epl/i2005-10085-3}}.
\newline\urlprefix\url{https://doi.org/10.1209/epl/i2005-10085-3}

\bibitem{bib:Efthymiopoulos2006}
C.~Efthymiopoulos, G.~Contopoulos,
  \href{https://doi.org/10.1088%2F0305-4470%2F39%2F8%2F004}{Chaos in bohmian
  quantum mechanics}, Journal of Physics A: Mathematical and General 39~(8)
  (2006) 1819--1852.
\newblock \href {http://dx.doi.org/10.1088/0305-4470/39/8/004}
  {\path{doi:10.1088/0305-4470/39/8/004}}.
\newline\urlprefix\url{https://doi.org/10.1088%2F0305-4470%2F39%2F8%2F004}

\bibitem{bib:Efthymiopoulos2007}
C.~Efthymiopoulos, C.~Kalapotharakos, G.~Contopoulos,
  \href{http://dx.doi.org/10.1088/1751-8113/40/43/008}{Nodal points and the
  transition from ordered to chaotic bohmian trajectories}, Journal of Physics
  A: Mathematical and Theoretical 40~(43) (2007) 12945–12972.
\newblock \href {http://dx.doi.org/10.1088/1751-8113/40/43/008}
  {\path{doi:10.1088/1751-8113/40/43/008}}.
\newline\urlprefix\url{http://dx.doi.org/10.1088/1751-8113/40/43/008}

\bibitem{bib:Efthymiopoulos2017}
C.~Efthymiopoulos, G.~Contopoulos, A.~C. Tzemos, Chaos in de broglie - bohm
  quantum mechanics and the dynamics of quantum relaxation, 2017.

\bibitem{bib:Ji1995}
J.-Y. Ji, J.~K. Kim, S.~P. Kim, K.-S. Soh, {Exact wave functions and
  nonadiabatic Berry phases of a time-dependent harmonic oscillator}, Phys.
  Rev. A52 (1995) 3352--3355.
\newblock \href {http://dx.doi.org/10.1103/PhysRevA.52.3352}
  {\path{doi:10.1103/PhysRevA.52.3352}}.

\bibitem{bib:Agarwal1991}
G.~S. Agarwal, S.~A. Kumar,
  \href{https://link.aps.org/doi/10.1103/PhysRevLett.67.3665}{Exact
  quantum-statistical dynamics of an oscillator with time-dependent frequency
  and generation of nonclassical states}, Phys. Rev. Lett. 67 (1991)
  3665--3668.
\newblock \href {http://dx.doi.org/10.1103/PhysRevLett.67.3665}
  {\path{doi:10.1103/PhysRevLett.67.3665}}.
\newline\urlprefix\url{https://link.aps.org/doi/10.1103/PhysRevLett.67.3665}

\bibitem{bib:Hairer1993}
E.~Hairer, S.~P. N\o{}rsett, G.~Wanner, Solving Ordinary Differential Equations
  I (2nd Revised. Ed.): Nonstiff Problems, Springer-Verlag, Berlin, Heidelberg,
  1993.

\bibitem{bib:Pinto-Neto2005}
N.~Pinto-Neto, \href{https://doi.org/10.1007/s10701-004-2012-8}{The bohm
  interpretation of quantum cosmology}, Foundations of Physics 35~(4) (2005)
  577--603.
\newblock \href {http://dx.doi.org/10.1007/s10701-004-2012-8}
  {\path{doi:10.1007/s10701-004-2012-8}}.
\newline\urlprefix\url{https://doi.org/10.1007/s10701-004-2012-8}

\bibitem{bib:Peter2008}
P.~Peter, N.~Pinto-Neto, {Cosmology without inflation}, Phys. Rev. D 78 (2008)
  063506.
\newblock \href {http://arxiv.org/abs/0809.2022} {\path{arXiv:0809.2022}},
  \href {http://dx.doi.org/10.1103/PhysRevD.78.063506}
  {\path{doi:10.1103/PhysRevD.78.063506}}.

\bibitem{bib:Pinto-Neto2012}
N.~Pinto-Neto, G.~Santos, W.~Struyve,
  \href{https://link.aps.org/doi/10.1103/PhysRevD.85.083506}{Quantum-to-classical
  transition of primordial cosmological perturbations in de broglie--bohm
  quantum theory}, Phys. Rev. D 85 (2012) 083506.
\newblock \href {http://dx.doi.org/10.1103/PhysRevD.85.083506}
  {\path{doi:10.1103/PhysRevD.85.083506}}.
\newline\urlprefix\url{https://link.aps.org/doi/10.1103/PhysRevD.85.083506}

\bibitem{bib:Pinto-Neto2013}
N.~Pinto-Neto, J.~C. Fabris, {Quantum cosmology from the de Broglie-Bohm
  perspective}, Class. Quant. Grav. 30 (2013) 143001.
\newblock \href {http://arxiv.org/abs/1306.0820} {\path{arXiv:1306.0820}},
  \href {http://dx.doi.org/10.1088/0264-9381/30/14/143001}
  {\path{doi:10.1088/0264-9381/30/14/143001}}.

\bibitem{bib:Pinto-Neto2016}
D.~C.~F. Celani, N.~Pinto-Neto, S.~D.~P. Vitenti, {Particle Creation in
  Bouncing Cosmologies}, Phys. Rev. D95~(2) (2017) 023523.
\newblock \href {http://arxiv.org/abs/1610.04933} {\path{arXiv:1610.04933}},
  \href {http://dx.doi.org/10.1103/PhysRevD.95.023523}
  {\path{doi:10.1103/PhysRevD.95.023523}}.

\bibitem{bib:Pinto-Neto2018}
N.~Pinto-Neto, W.~Struyve, {Bohmian quantum gravity and cosmology}\href
  {http://arxiv.org/abs/1801.03353} {\path{arXiv:1801.03353}}.

\bibitem{bib:Jerome2000}
J.~Martin, A.~Riazuelo, M.~Sakellariadou,
  \href{https://link.aps.org/doi/10.1103/PhysRevD.61.083518}{Nonvacuum initial
  states for cosmological perturbations of quantum-mechanical origin}, Phys.
  Rev. D 61 (2000) 083518.
\newblock \href {http://dx.doi.org/10.1103/PhysRevD.61.083518}
  {\path{doi:10.1103/PhysRevD.61.083518}}.
\newline\urlprefix\url{https://link.aps.org/doi/10.1103/PhysRevD.61.083518}

\bibitem{bib:Greene2006}
B.~R. Greene, M.~K. Parikh, J.~P. van~der Schaar,
  \href{https://doi.org/10.1088%2F1126-6708%2F2006%2F04%2F057}{Universal
  correction to the inflationary vacuum}, Journal of High Energy Physics
  2006~(04) (2006) 057--057.
\newblock \href {http://dx.doi.org/10.1088/1126-6708/2006/04/057}
  {\path{doi:10.1088/1126-6708/2006/04/057}}.
\newline\urlprefix\url{https://doi.org/10.1088%2F1126-6708%2F2006%2F04%2F057}

\end{thebibliography}
\bibliographystyle{ieeetr}

\end{document}